%% file: main.tex
\documentclass{article}
\usepackage{graphicx} 
\usepackage{authblk}
\usepackage{mathtools} 
\usepackage[utf8]{inputenc}
\usepackage[style=phys,sorting=none,backend=biber]{biblatex} 
\usepackage{amsmath,amssymb}
\usepackage{bm}
\usepackage{lmodern}
\usepackage[textwidth=16cm,textheight=23cm]{geometry}
\usepackage[hidelinks]{hyperref}
\hypersetup{colorlinks=false}

\usepackage[font=small, labelfont=bf]{caption}
\usepackage{subfig} 
\usepackage{float}

\usepackage{booktabs}   
\usepackage{array}
\usepackage[nameinlink, capitalise]{cleveref}

\usepackage{pgfplots}
\pgfplotsset{width=5cm,compat=1.9}

\usepgfplotslibrary{external}
\newcolumntype{x}[1]{%
>{\centering\hspace{0pt}}p{#1}}%

\usepackage{xr}
\makeatletter
\newcommand*{\addFileDependency}[1]{
  \typeout{(#1)}
  \@addtofilelist{#1}
  \IfFileExists{#1}{}{\typeout{No file #1.}}
}
\makeatother

\newcommand*{\myexternaldocument}[1]{%
    \externaldocument{#1}%
    \addFileDependency{#1.tex}%
    \addFileDependency{#1.aux}%
}
\myexternaldocument{SI}

\usepackage{enumitem}

\renewcommand{\vec}[1]{{\bf #1}} 

\usepackage[b]{esvect}
\usepackage{chngcntr}

\author[1]{Juliane H. Fuglsbjerg\footnote{Email: jf@chem.ku.dk}}
\author[3]{Sonia Coriani}
\author[2]{Jacob Kongsted}
\author[1]{Stephan P. A. Sauer}
\affil[1]{\textit{Department of Chemistry, University of Copenhagen, Universitetsparken 5, DK-2100 Copenhagen \O, Denmark.}} \affil[2]{\textit{Department of Physics, Chemistry and Pharmacy,
University of Southern Denmark, Campusvej 55, DK-5230 Odense M, Denmark.}}
\affil[3]{\textit{DTU Chemistry, Technical University of Denmark, Kemitorvet Building 207, DK-2800 Kongens Lyngby, Denmark.}}
\date{}
\title{Unitary Coupled Cluster Nuclear Shielding Constants in a Quantum Linear Response Framework}

\begin{document}
\maketitle

\begin{abstract} 
We present calculations of nuclear shielding constants, a key ingredient in nuclear magnetic resonance spectroscopy, within a quantum-computing-suitable framework.
We employ quantum linear response (qLR) theory in combination with the unitary coupled cluster (UCC) ansatz and its orbital-optimized (oo-UCC), factorized (fUCC), and orbital-optimized factorized (oo-fUCC) variants. The methods are assessed for seven small molecules against CASCI, CASSCF, CCSD, and CCSD(T) reference calculations. The different UCC-based qLR approaches yield shielding constants in good agreement with comparable classical methods. Factorization, which is necessary for quantum-computing implementations, has only a negligible effect on the calculated shielding constants. Orbital optimization, in contrast, leads to substantial improvements, with oo-UCC and oo-fUCC showing markedly better agreement with CCSD and CCSD(T). Our results establish oo-fUCC combined with qLR as a promising approach for computing nuclear shielding constants on quantum computers.
\end{abstract}

\section{Introduction}

Quantum computers are an emerging technology, promising solutions to chemically relevant but computationally intractable problems such as molecules with strong electron correlation. However, the current noisy intermediate-scale quantum (NISQ) devices are error-prone, necessitating the use of shallow quantum circuits and limited qubit counts, restricting calculations to systems with only a limited number of correlated electrons.
A well-known strategy to achieve these shallow circuit depths is to employ the complete active space approach well-known in traditional quantum chemistry \cite{siegbahn_comparison_1980, roos_complete_1980, siegbahn_complete_1981} also in quantum algorithms \cite{Fitzpatrick2024}.
To date, quantum-computing algorithms for NISQ devices have primarily targeted energies \cite{ollitrault_quantum_2020,asthana_quantum_2023,barison_quantum-centric_2025}, but computation of other molecular properties of spectroscopic interest is a growing field of research. Examples from the literature include oscillator strengths \cite{kumar_quantum_2023,ziems_which_2024,jensen_quantum_2024}, rotational strengths \cite{kumar_quantum_2023,von_buchwald_reduced_2024}, polarizabilities \cite{kumar_quantum_2023,reinholdt_subspace_2024}, hyperfine coupling constants \cite{jensen_hyperfine_2025} and spin-spin coupling constants \cite{fuglsbjerg_orbital-optimized_2026} of nuclear magnetic resonance spectroscopy (NMR). 

NMR spectroscopy is an important analytical technique in many areas of natural science.
The interpretation of modern NMR spectra increasingly involves the integration of experimental measurements with quantum-chemical calculations of chemical shifts and indirect nuclear spin-spin coupling constants \cite{Hansen2024Synergy, deAlbuquerque2024Reassignment, spas213}, making the quantum chemical determination of NMR properties an active field of research \cite{kaupp_calculation_2004, GasPhase}. The calculation of accurate shielding constants, especially non-hydrogen shieldings, requires (a) an adequate treatment of electron correlation, commonly coupled cluster singles doubles (CCSD) \cite{purvis_full_1982} with perturbative corrections for triple excitations (CCSD(T)) \cite{raghavachari_fifth-order_1989} and (b) a large basis set \cite{auer_quantitative_2003,harding_quantitative_2008,auer_quantitative_2009,kupka_convergence_2010}. 
CCSD(T), often regarded as the gold standard of quantum chemistry, exhibits a steep computational scaling with system size, with costs increasing approximately as $N^7$, where $N$ is a measure of the number of molecular orbitals. This unfavorable scaling severely restricts its application to relatively small molecular systems and often necessitates the use of less computationally demanding, but also less accurate, electronic structure methods for larger molecules. As a result, the accurate calculation of properties such as nuclear shielding constants becomes increasingly challenging as system size grows. These limitations make NMR shielding constants an attractive target for quantum computing applications, as they may enable the routine prediction of highly accurate shielding constants for molecular systems that are currently beyond the practical reach of classical CCSD(T) calculations.

In our recent work \cite{fuglsbjerg_orbital-optimized_2026}, we extended the quantum linear response (qLR) framework introduced by Ziems et al. \cite{ziems_which_2024} and Jensen et al. \cite{jensen_quantum_2024}, to enable the calculation of NMR spin-spin coupling constants on quantum computers.
Here, we apply this qLR framework to calculate NMR shielding constants in a quantum-computing-suitable manner, to our knowledge the first of such applications.
The qLR formalism is combined with the unitary coupled cluster (UCC) ansatz \cite{bartlett_alternative_1989,peruzzo_variational_2014} and its orbital-optimized variant (oo-UCC) \cite{mizukami_orbital_2020,sokolov_quantum_2020,bierman_improving_2023}, both of which are particularly well suited for quantum computing due to their unitary structure and factorization properties, enabling efficient implementation in terms of quantum gates.

Results are presented for the NMR shielding constants of seven small molecules in a variety of active spaces. 
To assess the quality of the UCC shielding constants, they are compared to shielding constants calculated with the classical methods: complete active space self consistent field (CASSCF) \cite{hinze_multi-configuration_1967,adams_orbital_1967} and complete active space configuration interaction (CASCI) \cite{Levine2021CASCI}, as well as conventional full-space CCSD and CCSD(T). Four different unitary coupled cluster ans{\"a}tze are tested to investigate both the impact of including orbital rotations in the wavefunction and linear response, and the impact of factorizing the wavefunction, a necessary step when encoding the wavefunction on quantum hardware.

\section{Theory}

\subsection{Nuclear Shielding Constants} \label{sec:shield}

In NMR spectroscopy, each nucleus $A$, is shielded from the full external magnetic field, $\vec{B}$, by the surrounding electrons in the molecule. Nucleus $A$ therefore interacts with a local magnetic field \cite{fukui_theory_1997,kaupp_calculation_2004}
\begin{equation}
    \vec{B}_{A,loc} = (1-{\boldsymbol{\sigma}}_A)\vec{B} \,,
\end{equation}
where $\boldsymbol{\sigma}_A$ is the nuclear shielding tensor of $A$ for the specific environment. Any nucleus with an associated nuclear magnetic dipole moment, $\vec{m}_A$, will interact with the local magnetic field, resulting in an interaction energy
\begin{equation}
    E_A = -\vec{m}_A \cdot \vec{B}_{A,loc} = -\vec{m}_A \cdot (1-\boldsymbol{\sigma}_A)\vec{B} \,.
\end{equation}
The shielding tensor can be related to the chemical shift observed in NMR spectroscopy, but in liquid and gas-phase NMR, only one component of the shielding tensor is observed due to the free rotation of molecules. This component is known as the isotropic shielding constant
\begin{equation}
    \sigma_{A} = 
    \frac{1}{3} \sum_\mu \sigma_{A,\mu\mu} \,.
\end{equation}
where the Greek subscript $\mu$ (and later $\nu$) denotes a Cartesian component, i.e. $x$, $y$ or $z$.
To calculate the isotropic shielding constant, the necessary components can be retrieved as derivatives of the perturbed electronic energy with respect to the components of the nuclear magnetic moment and the external magnetic field
\begin{equation} \label{eq:shield_gen}
    \sigma_{A,\mu\mu} = \left. \frac{\partial^2 E}{\partial m_{A,\mu}\partial B_{\mu}} \right|_{\vec{m}_A=\vec{B}=\vec{0}} \,.
\end{equation}
Only the second-order correction to the energy contributes to the shielding \cite{nmr500715r}, which is given by
\begin{equation} \label{eq:2nd_E}
\begin{split}
    E^{(2)}(\vec{m}_A,\vec{B}) =& \left\langle 0 \left| \hat{H}^{(2)}(\vec{m}_A,\vec{B}) \right| 0 \right\rangle  + \sum_{n\neq 0} \frac{\langle 0 | \hat{H}^{(1)}(\vec{m}_A, \vec{B}) | n\rangle\langle n | \hat{H}^{(1)}(\vec{m}_A,\vec{B}) | 0\rangle}{E_0 - E_n} \,,
\end{split}
\end{equation}
where $|0\rangle$ is the unperturbed ground state, $E_0$ is the ground state energy, $|n\rangle$ is the unperturbed $n$'th excited state and $E_n$ is its corresponding energy, and $\hat{H}^{(1)}$ and $\hat{H}^{(2)}$ are the first- and second-order corrections to the unperturbed Hamiltonian.

In the minimal coupling approximation, the magnetic fields and moments enter the Hamiltonian as perturbations in the form of the vector potential, $\hat{\vec{A}}(\vec{r})$ \cite{spasB8}. Assuming the Coulomb gauge, $\boldsymbol{\nabla} \cdot \vec{A} = 0$, the first-order correction to the spin-free Hamiltonian in second-quantization notation is 
\begin{equation} 
\label{eq:H1_gen}
    \hat{H}^{(1)} = \sum_{pq} h^{(1)}_{pq} \hat{E}_{pq} \,,
\end{equation}
where $p$ and $q$ are general spatial orbital indices and $h^{(1)}_{pq}$ is a 
one-electron integral defined as
\begin{equation}
    h^{(1)}_{pq} = \int \phi_p^*(\vec{r}) \left( - i \hat{\vec{A}}(\vec{r}) \cdot \hat{\boldsymbol\nabla} \right) \phi_q(\vec{r}) \mathrm{d}\vec{r} \,,
\end{equation}
where $\vec{r}$ is the position vector of the electron, $\phi_p$ is the $p$'th molecular orbital and $\hat{E}_{pq}$ is the singlet one-electron excitation operator from orbital $q$ to orbital $p$
\begin{equation}
    \hat{E}_{pq} =~ \hat{a}^\dagger_{p\alpha} \hat{a}_{q\alpha} + \hat{a}_{p\beta}^\dagger \hat{a}_{q\beta} \,.
\end{equation}
The second-order correction to the Hamiltonian is
\begin{equation} \label{eq:H2_gen}
    \hat{H}^{(2)} = \frac{1}{2} \sum_{pq} h^{(2)}_{pq} \hat{E}_{pq} \,,
\end{equation}
where $h^{(2)}_{pq}$ is defined as
\begin{equation}
    h^{(2)}_{pq} = \int \phi_p^*(\vec{r}) \hat{\vec{A}}^2(\vec{r}) \phi_q(\vec{r}) \mathrm{d}\vec{r} \,.
\end{equation}
The vector potential arises from both the nuclear magnetic moment of nucleus $A$ and the external magnetic field, resulting in two contributions to the vector potential
\begin{align}
    \hat{\vec{A}} (\vec{r}) =~& \frac{1}{2} \vec{B} \times (\vec{r} - \vec{R}_{GO}) +  \vec{m}_A \times \frac{(\vec{r}-\vec{R}_A)}{|\vec{r} - \vec{R}_A|^3} \,,
\end{align}
where $\vec{R}_{GO}$ is the position vector of the gauge-origin and $\vec{R}_A$ is the position vector of $A$. The dependence on the gauge-origin arises because the vector potential is not uniquely defined from the external magnetic field, allowing for the transformation of the vector potential e.g. by moving the gauge-origin, without changing the external magnetic field. In a complete basis, a gauge-transformation of the vector potential has no impact on properties, but that does not hold for a finite basis, introducing an unphysical dependence on the choice of gauge-origin \cite{fukui_theory_1997}. 

By inserting the vector potentials in the first and second order corrections in \cref{eq:H1_gen,eq:H2_gen}, expressions for the Hamiltonian corrections can be derived. 
Retaining only the terms relevant for the calculation of the shielding constants
yields
\begin{equation} \label{eq:H_1st}
    \hat{H}_A^{(1)} = \sum_\mu \left(\hat{O}^{OP}_{A,\mu} m_{A,\mu} + \hat{O}^{l}_\mu B_{\mu} \right) \,,
\end{equation} 
\begin{equation} \label{eq:H_2nd}
    \hat{H}_A^{(2)} = \sum_{\mu\nu} \hat{O}_{A,\mu\nu} m_{A,\mu} B_\nu \,.
\end{equation}
The individual perturbation operators in the Hamiltonian are defined as
\begin{align}
    \hat{O}^{OP}_{A,\mu} ~=~& 
    \sum_{pq} 
    \int \phi_p^*(\vec{r}) 
    \frac{\hat{l}_\mu(\vec{R}_A)}{|\vec{r} - \vec{R}_A|^3} 
     \phi_q(\vec{r}) \mathrm{d}\vec{r} \,
    \hat{E}_{pq} \label{eq:op} \,, \\
    \hat{O}^{l}_\mu ~=~& 
    \frac{1}{2} 
    \sum_{pq}
    \int \phi_p^*(\vec{r})
    \hat{l}_{\mu}(\vec{R}_{GO})
    \phi_q(\vec{r}) \mathrm{d}\vec{r}
    \, \hat{E}_{pq} \label{eq:op_l} \,, \\
    \hat{O}^{DS}_{A,\mu\nu}~=~&
    \frac{1}{2}
    \sum_{pq}
    \int \phi_p^*(\vec{r})
    \left(
    (\vec{r}-\vec{R}_{GO})\cdot\frac{(\vec{r}-\vec{R}_A)}{|\vec{r}-\vec{R}_A|^3}\delta_{\mu\nu}
    - (r_\mu - R_{GO,\mu})\frac{(r_\nu - R_{A,\nu})}{|\vec{r}-\vec{R}_A|^3}
    \right)
    \phi_q(\vec{r}) \mathrm{d}\vec{r}
    \, \hat{E}_{pq} \,,
\end{align}
where $\hat{{l}}_\mu$ is the $\mu$ Cartesian component of the orbital angular momentum operator. $\hat{O}^{OP}_{A,\mu}$ is the orbital paramagnetic (OP) operator, $\hat{O}^{l}_\mu$ is the orbital magnetic dipole operator and $\hat{O}_{A,\mu\nu}$ is the diamagnetic shielding (DS) operator. Inserting the first and second order corrections from \cref{eq:H_1st,eq:H_2nd} in \cref{eq:2nd_E} for the energy yields
\begin{equation}
\label{the2ndorderE}
\begin{split}
    E^{(2)}(\vec{m}_A,\vec{B}) = &
    \sum_{\mu\nu} m_{A,\mu} B_{\nu} \left\langle 0 \left|  \hat{O}^{DS}_{A,\mu\nu}  \right| 0 \right\rangle  \\
    &+ 
    \sum_{\mu\nu} \sum_{n\neq 0} \frac{\langle 0 | m_{A,\mu} \hat{O}^{\mathrm{OP}}_{A,\mu} + B_{\mu}\hat{O}^l_\mu | n\rangle\langle n | m_{A,\nu} \hat{O}^{\mathrm{OP}}_{A,\nu} + B_{\nu}\hat{O}_\nu^l | 0\rangle}{E_0 - E_n} \,.
\end{split}
\end{equation}
Inserting the energy expression in \cref{the2ndorderE}
into \cref{eq:shield_gen} to calculate the components of the shielding tensor gives
\begin{equation}
\begin{split}
    \sigma_{A,\mu\mu} =& \left. \frac{\partial^2 E^{(2)}(\vec{m}_A,\vec{B})}{\partial m_{A,\mu}\partial B_{\mu}} \right|_{\vec{m}_A=\vec{B}=0} \\
    =& \left\langle 0 \left| \hat{O}^{DS}_{A,\mu\mu} \right| 0 \right\rangle 
    + 
    \sum_{n\neq 0} \left( \frac{\langle 0 | \hat{O}^{\mathrm{OP}}_{A,\mu} | n\rangle\langle n | \hat{O}_\mu^l | 0\rangle}{E_0 - E_n} 
    +
    \frac{\langle 0 | \hat{O}^l_\mu | n\rangle\langle n | \hat{O}^{\mathrm{OP}}_{A,\mu} | 0\rangle}{E_0 - E_n} \right)
    \\
    =& \left\langle 0 \left| \hat{O}^{DS}_{A,\mu\mu} \right| 0 \right\rangle + \left\langle\left\langle \hat{O}^{OP}_{A,\mu};\hat{O}^l_{\mu} \right\rangle \right\rangle \,,
\end{split}
\end{equation}
where $\left\langle\left\langle \hat{O}^{OP}_{A,\mu};\hat{O}^l_{\mu} \right\rangle \right\rangle$ is a (frequency-independent) linear response function. 
The first term is the diamagnetic contribution to the nuclear shielding constant,
and it is computed as ground state expectation value of the diamagnetic shielding operator.
The second term is the paramagnetic contribution, which is computed as the response function detailed in \cref{sec:qLR}.
Note that both contributions to the shielding constant depend on the gauge-origin $\vec{R}_{GO}$ through $\hat{O}^{DS}_{A,\mu\mu}$ and $\hat{O}^{l}_{\mu}$, respectively.

\subsection{Unitary Coupled Cluster Wavefunction} \label{sec:ucc}

A well-known ansatz for quantum computing is the chemically-inspired UCC ansatz \cite{bartlett_alternative_1989,peruzzo_variational_2014} and its orbital-optimized variant (oo-UCC) \cite{mizukami_orbital_2020,sokolov_quantum_2020,bierman_improving_2023}. UCC is defined as an exponentiated anti-hermitian operator acting on the Hartree-Fock state
\begin{equation} \label{eq:ucc}
    |\text{UCC}(\boldsymbol{\theta})\rangle = \exp \left\{ {\sum_n \theta_n \left( \hat{G}_n - \hat{G}^\dagger_n \right)} \right\} |\mathrm{HF} \rangle \,,
\end{equation}
where $\hat{G}_n$ is, for UCC singles and doubles (UCCSD), one of the following single or double (spin-adapted) excitation operators
\begin{align}
    \hat{G}'_n =& \,\frac{1}{\sqrt{2}} \hat{E}_{v_a v_i} \,, \label{eq:G1} \\
    \hat{G}''_n =& \,
    \frac{1}{2}\frac{1}{\sqrt{(1+\delta_{v_a v_b})(1+\delta_{v_i v_j})}} 
    (\hat{E}_{v_a v_i}\hat{E}_{v_b v_j} + \hat{E}_{v_a v_j}\hat{E}_{v_b v_i}) \,, \label{eq:G2} \\
    \hat{G}'''_n =& \,
    \frac{1}{2\sqrt{3}}
    (\hat{E}_{v_a v_i}\hat{E}_{v_b v_j} - \hat{E}_{v_a v_j}\hat{E}_{v_b v_i}) \,, \label{eq:G3}
\end{align}
where $v$ is an active orbital index, meaning that the ansatz is only applied within the active space. The subscript denotes whether the orbital is occupied, $i$, $j$, or unoccupied, $a$, $b$, in the Hartree-Fock wavefunction. 
To implement the UCCSD wavefunction on quantum hardware, it is necessary to factorize it, i.e., letting the ansatz be a product of individual exponentiated operators as opposed to the sum in \cref{eq:ucc}. This results in a factorized UCCSD (fUCCSD), which is an approximation to the non-factorized UCCSD
\begin{equation}
    |\text{fUCC}(\boldsymbol{\theta})\rangle = \prod_n \exp \left\{ {\theta_n\left( \hat{G}_n - \hat{G}^\dagger_n \right)} \right\} |\mathrm{HF} \rangle \,.
\end{equation}
Orbital-optimized UCCSD (oo-UCCSD) and fUCCSD (oo-fUCCSD) are obtained by letting an orbital rotation operator act on the UCCSD and fUCCSD wavefunctions
\begin{equation}
    |\text{oo-UCC}(\boldsymbol{\theta},\boldsymbol{\kappa})\rangle = 
    \exp \left\{ {-\hat{\kappa}(\boldsymbol{\kappa})} \right\} |\mathrm{UCC}(\boldsymbol{\theta})\rangle \,,
\end{equation}
\begin{equation}
    |\text{oo-fUCC}(\boldsymbol{\theta},\boldsymbol{\kappa})\rangle = 
    \exp \left\{ {-\hat{\kappa}(\boldsymbol{\kappa})} \right\} |\mathrm{fUCC}(\boldsymbol{\theta})\rangle \,,
\end{equation}
where $\hat{\kappa}(\boldsymbol{\kappa})$ is the orbital rotation operator defined as
\begin{equation}
    \hat{\kappa}(\boldsymbol{\kappa}) = \sum_{vi} \kappa^v_i \left( \hat{E}_{vi} - \hat{E}_{iv}\right)
    +
    \sum_{ai} \kappa^a_i \left( \hat{E}_{ai} - \hat{E}_{ia}\right)
    +
    \sum_{av} \kappa^a_v \left( \hat{E}_{av} - \hat{E}_{va}\right) \,.
\end{equation}
The indices are $i$ for inactive orbitals, $v$ for active orbitals and $a$ for virtual orbitals. The restriction on the indices ensures that only non-redundant orbital rotations---i.e., rotations between the spaces---are included.

The parameters are obtained through variational minimization of the (unperturbed) ground state energy. The energies are evaluated on quantum hardware and the parameters are then optimized classically. In the case of oo-UCCSD and oo-fUCCSD, it is convenient to let the orbital rotation operator act on the Hamiltonian to avoid explicit manipulation of the inactive and virtual orbitals on the quantum device
\begin{align}
    E_\text{UCC} =& \min_{\boldsymbol{\theta}}\langle \text{UCC}(\boldsymbol{\theta}) | \hat{H} | \text{UCC}(\boldsymbol{\theta}) \rangle \,, \\
    E_\text{fUCC} =& \min_{\boldsymbol{\theta}}\langle \text{fUCC}(\boldsymbol{\theta}) | \hat{H} | \text{fUCC}(\boldsymbol{\theta}) \rangle \,, \\
    E_\text{oo-UCC} =& \min_{\boldsymbol{\theta},\boldsymbol{\kappa}}\langle \text{UCC}(\boldsymbol{\theta}) | \hat{H}(\boldsymbol{\kappa}) | \text{UCC}(\boldsymbol{\theta}) \rangle \,, \\
    E_\text{oo-fUCC} =& \min_{\boldsymbol{\theta},\boldsymbol{\kappa}}\langle \text{fUCC}(\boldsymbol{\theta}) | \hat{H}(\boldsymbol{\kappa}) | \text{fUCC}(\boldsymbol{\theta}) \rangle \,,
\end{align}
where the rotated Hamiltonian is
\begin{equation}
    \hat{H}(\boldsymbol{\kappa}) = \exp \left\{ {\hat{\kappa}(\boldsymbol{\kappa})} \right\}  \hat{H} \exp \left\{ {-\hat{\kappa}(\boldsymbol{\kappa})} \right\} \,.
\end{equation}
All four ans{\"a}tze will be investigated for the calculation of NMR shielding constants in \cref{sec:results}.

\subsection{Quantum linear response for NMR shieldings} \label{sec:qLR}

Quantum linear response theory is adopted from the naive parametrization introduced by Ziems et al. \cite{ziems_which_2024}, which is based on classical MCSCF linear response theory \cite{olsen_linear_1985,helgaker_recent_2012}. The necessary equations for the calculation of NMR shielding constants are summarized here for the convenience of the reader. The (frequency-independent) response function for the operators $\hat{O}$ and $\hat{P}$ is given as
\begin{equation} \label{eq:response_function}
    \langle \langle \hat{O} ; \hat{P} \rangle \rangle = -{\mathbf{V}_O^{[1]}}^* \cdot \boldsymbol{\beta}_P \,,
\end{equation}
where $\mathbf{V}_O^{[1]}$ is the property gradient vector of $\hat{O}$ and $\boldsymbol{\beta}_P$ is the response vector of $\hat{P}$.
The response vector is found through solving the response equation
\begin{equation} \label{eq:linear_response}
    \mathbf{E}^{[2]} \boldsymbol{\beta}_P = \mathbf{V}_P^{[1]} \,,
\end{equation}
where $\mathbf{E}^{[2]}$ is the electronic Hessian. The property gradient and electronic Hessian are defined as follows
\begin{equation}
    \mathbf{V}_P^{[1]} = 
    \begin{pmatrix}
        \mathbf{X}_{P}
        \\ -\mathbf{X}_{P}^*
    \end{pmatrix} \,,
\end{equation}
\begin{equation}
    \mathbf{X}_P = 
    \begin{pmatrix}
        \langle 0 | [\hat{q}_m, \hat{P}] | 0 \rangle
        \\ \langle 0 | [\hat{G}_n, \hat{P}] | 0 \rangle
    \end{pmatrix} \,,
\end{equation}
\begin{equation}
    \mathbf{E}^{[2]} = 
    \begin{pmatrix}
        \mathbf{A} & \mathbf{B} \\ \mathbf{B}^* & \mathbf{A}^* 
    \end{pmatrix} \,,
\end{equation}
\begin{equation}
    \mathbf{A} = 
    \begin{pmatrix}
        \langle  0 | [\hat{q}^\dagger_m, \hat{H}, \hat{q}_{m '}] | 0 \rangle
        & \langle  0 | [\hat{q}^\dagger_{m}, \hat{H}, \hat{G}_{n'}] | 0 \rangle
        \\ \langle  0 | [\hat{G}^\dagger_{n}, \hat{H}, \hat{q}_{m '}] | 0 \rangle
        & \langle  0 | [\hat{G}^\dagger_n, \hat{H}, \hat{G}_{n'}] | 0 \rangle
    \end{pmatrix} \,,
\end{equation}
\begin{equation}
    \mathbf{B} = 
    \begin{pmatrix}
        \langle  0 | [\hat{q}^\dagger_m, \hat{H}, \hat{q}_{m '}^{\dagger}] | 0 \rangle
        & \langle  0 | [\hat{q}^\dagger_{m}, \hat{H}, \hat{G}_{n'}^{\dagger}] | 0 \rangle
        \\ \langle  0 | [\hat{G}^\dagger_{n}, \hat{H}, \hat{q}_{m '}^{\dagger}] | 0 \rangle
        & \langle  0 | [\hat{G}^\dagger_n, \hat{H}, \hat{G}_{n'}^{\dagger}] | 0 \rangle
    \end{pmatrix} \,,
\end{equation}
where $\hat{G}_n$ is defined in \cref{eq:G1,eq:G2,eq:G3}, $\hat{q}_m$ is one of the following (spin-adapted) orbital rotation operators
\begin{align}
    \hat{q}_m'   =&~ \frac{1}{\sqrt{2}} \hat{E}_{vi} \,, \\
    \hat{q}_m''  =&~ \frac{1}{\sqrt{2}} \hat{E}_{ai} \,, \\
    \hat{q}_m''' =&~ \frac{1}{\sqrt{2}} \hat{E}_{av} \,,
\end{align}
and the symmetrized double commutator is defined as
\begin{equation}
    [\hat{G}, \hat{H}, \hat{q}] = \frac{1}{2} \left( [\hat{G}, [\hat{H}, \hat{q}]] + [\hat{q}, [\hat{H}, \hat{G}]] \right) \,.
\end{equation}
For calculation of the paramagnetic contribution to NMR shielding constants, the operators $\hat{O}$ and $\hat{P}$ in the response function are the $\hat{O}^{l}_{\mu}$ and $\hat{O}^{OP}_{A,\mu}$ operators in \cref{eq:op,eq:op_l}.

\section{Computational details} \label{comp_det}

The seven molecules investigated are: H$_2$O, CH$_4$, NH$_3$, CO, HF, HCN, and N$_2$. All molecules are closed-shell systems.
The molecular geometries were optimized with second-order M{\o}ller-Plesset perturbation theory (MP2) \cite{moller_note_1934} in the 6-31G basis set \cite{ditchfield_selfconsistent_1971,hehre_selfconsistent_1972} using Gaussian16, revision A.03 \cite{gausian03}.

The NMR shielding constants were calculated with the four UCC methods introduced in \cref{sec:ucc}: UCCSD, fUCCSD, oo-UCCSD, and oo-fUCCSD, within the active space approximation. From here on, the four approaches will collectively be referred to as the UCC methods.

The active spaces utilized are denoted ($n_e$, $n_o$) where $n_e$ is the (even) number of active electrons and $n_o$ is the number of active spatial orbitals. The orbitals in the active space are the $n_e/2$ highest occupied orbitals and the $n_o-n_e/2$ lowest unoccupied orbitals. All inactive orbitals are doubly occupied, and all virtual orbitals are unoccupied. The chosen active spaces vary in size from (4,4) to (8,8). These represent spaces that are realistic on noisy intermediate scale quantum (NISQ) devices, as they have appropriate qubit counts and circuit depth. The exact active spaces for each molecule are listed in the tabulated results in \cref{sec:results_si} in the Supporting Information.

The UCC calculations are carried out using our in-house software package SlowQuant \cite{kjellgren_2026_22881628}, which is interfaced with PySCF version 2.7.0 \cite{sun_libcint_2015,sun_recent_2020} for the initial Hartree-Fock orbitals and for diamagnetic shielding, orbital paramagnetic, and orbital magnetic dipole integrals. In the case of oo-UCCSD and oo-fUCCSD on H$_2$O(6,5), PySCF was also utilized for calculating initial CASSCF orbitals to start the optimization from, since the calculations were otherwise not converging to the ground state. The subsequent linear response calculations are also carried out in SlowQuant with the naive linear response parametrization \cite{ziems_which_2024}. For the UCCSD and fUCCSD linear response calculations, we disabled orbital optimization in the linear response code in SlowQuant/slowquant/unitary\_coupled\_cluster/linear\_response/lr\_baseclass.py.

The basis set employed in the calculations of the NMR shielding calculations is the 6-31G** basis set \cite{ditchfield_selfconsistent_1971,hehre_selfconsistent_1972,hariharan_influence_1973} from basis set exchange \cite{schuchardt_basis_2007,pritchard_new_2019}. 

For comparison the UCC and oo-UCC results are benchmarked against CASCI and CASSCF, respectively, utilizing the same active spaces. The CASCI and CASSCF calculations are carried out using the Dalton2024.0 program \cite{aidas_dalton_2014, noauthor_dalton_2022}. To asses the relative quality of the various unitary coupled cluster results, the methods are benchmarked against conventional full-space unrelaxed CCSD. To investigate the effect of triples, the methods are also compared against unrelaxed CCSD(T). CCSD and CCSD(T) are carried out using the CFOUR program, version 1.2 \cite{matthews_coupled-cluster_2020}.

For all shielding constants and all computational methods, the calculations are carried out with the gauge-origin placed on the nuclei corresponding to the shielding constant, which allows for direct comparison between methods.

\section{Results and discussion} \label{sec:results}

The tabulated results from the oo-UCCSD, oo-fUCCSD, UCCSD, fUCCSD, CASSCF, CASCI, CCSD and CCSD(T) shielding constant calculations can be found in \cref{sec:results_si} in the Supporting Information. There are in total 14 calculated shieldings for the full-space methods, CCSD and CCSD(T), and 34 calculated shieldings for the active space methods.
In the following, we will compare the results of the different methods in terms of relative deviations, 
mean deviation (MD), mean absolute deviation (MAD), and standard deviation (SD) from either the CASSCF/CASCI or the CCSD and CCSD(T) reference results.

\subsection{Deviations from CASSCF and CASCI}

\begin{figure}[h!]
    \centering
    \subfloat[with orbital rotations]{\includegraphics[width=0.4\linewidth]{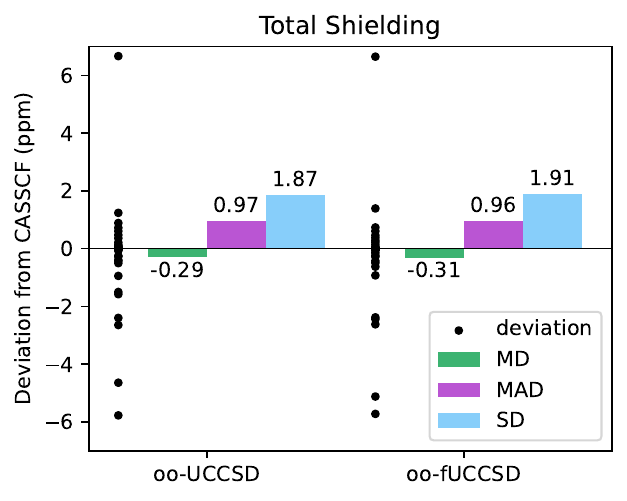}\label{fig:total_casscf}}
    \subfloat[without orbital rotations]{\includegraphics[width=0.4\linewidth]{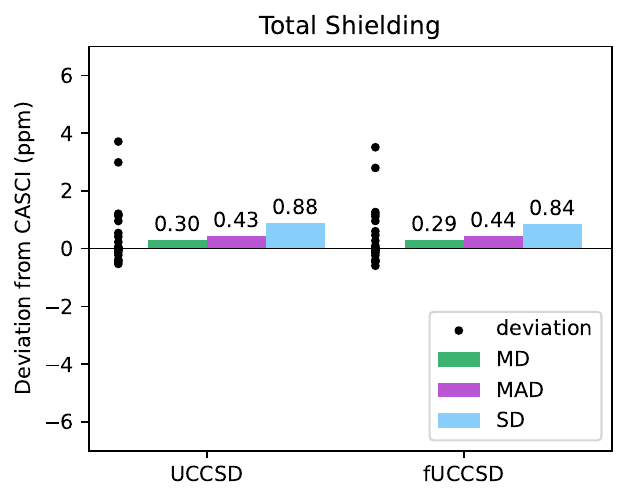}\label{fig:total_casci}}
    \caption{Total shielding deviations, MD, MAD and SD of \textbf{(a)} oo-UCCSD and oo-fUCCSD from CASSCF and \textbf{(b)} UCCSD and fUCCSD from CASCI. The 34 points depict the individual deviations for every shielding in every molecule and active space. The bars depict the MD (green), MAD (purple), and SD (blue).}
    \label{fig:total_both}
\end{figure}

The deviations, MD, MAD, and SD between CASSCF and oo-UCCSD/oo-fUCCSD and between CASCI and UCCSD/fUCCSD are depicted in \cref{fig:total_both}.

The deviations from CASSCF/CASCI arise because of the truncation in the UCC methods after double excitations, whereas the CASSCF/CASCI wavefunctions include all excitations within the active space. Overall the UCC methods are in good agreement with their classical counterparts, which can be seen from the deviations in \cref{fig:total_both}. oo-UCCSD and oo-fUCCSD both have a MAD of less than 1 ppm from CASSCF ($0.97$ and $0.96$ ppm, respectively). Similarly, UCCSD and fUCCSD have a MAD of less than 0.5 ppm from CASCI ($0.43$ and $0.44$ ppm, respectively). However all four UCC methods have a few outliers where the agreement with CASSCF/CASCI is not as good.

For oo-UCCSD the outliers that deviate from CASSCF with more than $2$ ppm are: $\sigma_\mathrm{O}$ in CO(8,7) ($-4.64$ ppm), $\sigma_\mathrm{N}$ in HCN(6,6) ($-2.40$ ppm), $\sigma_\mathrm{N}$ in HCN(8,8) ($6.68$ ppm), $\sigma_\mathrm{N}$ in N$_2$(6,6) ($-2.65$ ppm), and $\sigma_\mathrm{N}$ in N$_2$(8,8) ($-5.78$ ppm). All five outliers are in triple bonded molecules where higher excitations in the active space tend to be important. Three out of five are among the largest active spaces investigated. Generally, we would expect the deviation between oo-UCCSD/oo-fUCCSD and CASSCF to grow with increasing active space size, as more excitations are getting excluded by the truncation in oo-UCCSD/oo-fUCCSD. The same five shielding constants are also the outliers for oo-fUCCSD.

For UCCSD the shieldings that deviate from CASCI with more than $2$ ppm are: $\sigma_\mathrm{O}$ in CO(6,5) ($2.99$ ppm) and $\sigma_\mathrm{O}$ in CO(6,6) ($3.71$ ppm). Notably, for the largest active space, CO(8,7), the deviation from CASCI is much smaller ($0.03$ ppm), which is the opposite of the expected behavior and likely due to error cancellation. The same two outliers are also found for fUCCSD. 

The triple bonded molecules, CO, HCN, and N$_2$, consequently deviate more than the other molecules from CASSCF/CASCI, indicating the need for higher excitations in the active space for these systems. This is in line with the literature, as triple bonds are known to exhibit strong correlation \cite{Ruud1994, gauss_perturbative_1996}.  
The remainder of the molecules are well described despite the truncation after double excitations.

The performance of the factorized methods, oo-fUCCSD and fUCCSD, is nearly indistinguishable from the performance of the corresponding non-factorized methods, oo-UCCSD and UCCSD. This can be highlighted by, e.g., the MAD of the different methods. The MAD of oo-UCCSD ($0.97$ ppm) and oo-fUCCSD ($0.96$ ppm) are nearly identical to each other. The MAD of UCCSD ($0.43$ ppm) and fUCCSD ($0.44$ ppm) are also nearly identical to each other. From this we conclude that the factorization of the UCC wavefunction has minimal impact on the quality of the resulting shielding constants.

On the other hand, the deviation of the orbital optimized methods, 
oo-(f)UCCSD, from CASSCF and that of the non-optimized ones, (f)UCCSD, from CASCI are not the same. Comparison of the MAD, and SD shows that the MAD and SD of oo-(f)UCCSD is about twice as large as those for (f)UCCSD. 
This indicates that the higher excitations within the active space are more important when the wavefunction includes orbital rotations, as the deviations for those methods are larger.

The shielding constant can be separated in two contributions, the diamagnetic and paramagnetic contribution. To investigate the source of the deviations, the plots have been separated in a diamagnetic and paramagnetic contribution in \cref{fig:dia_both,fig:para_both} in the Supporting Information. From the figures it is evident that the source of the deviation between CASSCF/CASCI and the UCC methods arise exclusively from the paramagnetic contribution. The diamagnetic contribution is equally well described by all methods.

\subsection{Deviation from CCSD}

\begin{figure}[h!]
    \centering
    \subfloat[all active spaces]{\includegraphics[width=0.85\linewidth]{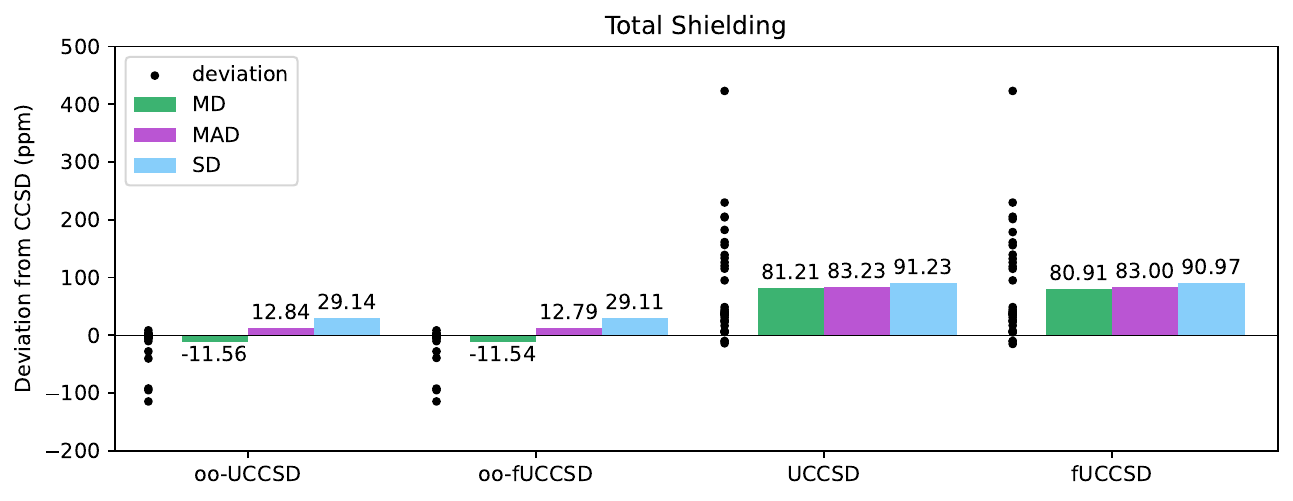}\label{fig:total_ccsd_all}}
    
    \subfloat[largest active spaces]{\includegraphics[width=0.85\linewidth]{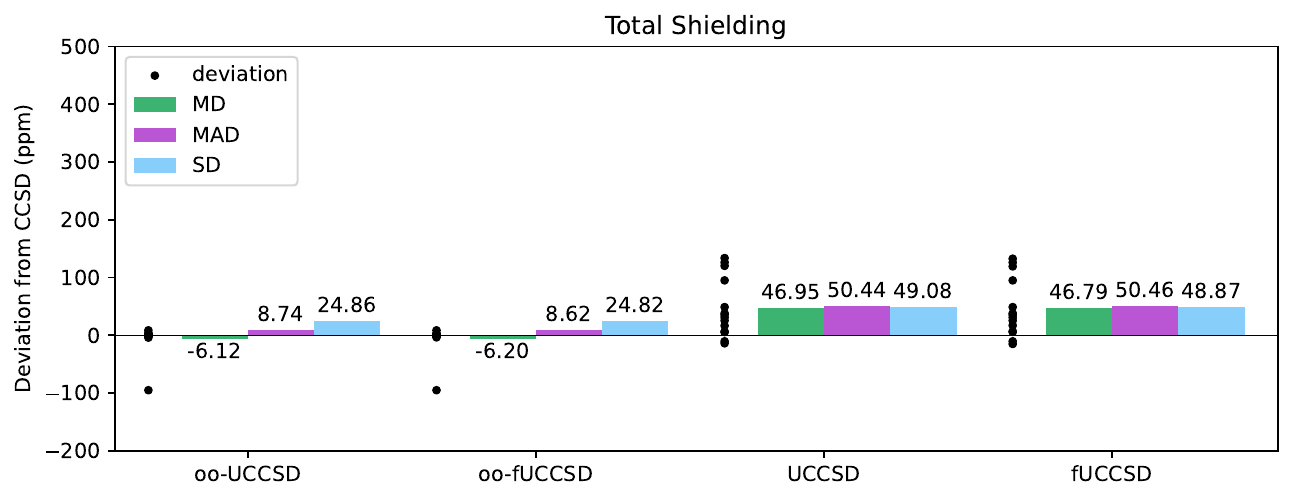}\label{fig:total_ccsd_best}}
    \caption{Total shielding deviations, MD, MAD and SD of oo-UCCSD, oo-fUCCSD, UCCSD and fUCCSD from (full-space) CCSD. The points depict the individual deviations for every shielding in every molecule for \textbf{(a)} every active space (34 points) and \textbf{(b)} the largest active space (14 points). The bars depict the MD (green), MAD (purple), and SD (blue).}
    \label{fig:total_ccsd}
\end{figure}

The results of the shielding constants calculated with and without orbital optimization can vary quite drastically, see for example $\sigma_O$ in CO in \cref{tab:co_o}, where oo-UCCSD(8,8) gives $-36.49$ ppm and UCCSD(8,8) gives $93.74$ ppm. It is therefore necessary to asses which of the methods result in more accurate shielding constants. All four UCC methods are compared to (full-space) conventional CCSD, to investigate which UCC methods is in better agreement with CCSD. The deviations, MD, MAD and SD between the UCC methods and CCSD are depicted in \cref{fig:total_ccsd_all}.

The addition of orbital optimization has a large impact on the shielding constants as can be seen in \cref{fig:total_ccsd_all}, where the deviations for the oo-UCCSD and oo-fUCCSD are significantly smaller than the deviations of UCCSD and fUCCSD. Comparing the MAD from CCSD of the four UCC methods, it is evident that oo-UCCSD ($12.84$ ppm) and oo-fUCCSD ($12.79$ ppm) are in far better agreement with CCSD than UCCSD ($83.23$ ppm) and fUCCSD ($83.00$ ppm). 

Notable outliers for oo-UCCSD, with an absolute deviation from CCSD larger than $90$ ppm, are: $\sigma_\mathrm{N}$ in HCN(4,4) ($-114.62$ ppm), $\sigma_\mathrm{N}$ in HCN(6,6) ($-92.20$ ppm) and $\sigma_\mathrm{N}$ in HCN(8,8) ($-95.16$ ppm). The same three shieldings are outliers for oo-fUCCSD.
For UCCSD the outliers with an absolute deviation from CCSD larger than $200$ ppm are: $\sigma_\mathrm{O}$ in CO(6,5) ($204.17$ ppm), $\sigma_\mathrm{C}$ in HCN(4,4) ($205.25$ ppm), $\sigma_\mathrm{N}$ in N$_2$(4,4) ($423.02$ ppm). fUCCSD has the same three shieldings as outliers.

Many of the large deviations from CCSD arise from the use of small active spaces. To investigate how choosing a larger active space impacts the quality of the results of the UCC methods, the deviations of all four UCC methods from (full-space) CCSD are plotted using only the largest active space employed for each molecule in \cref{fig:total_ccsd_best}.

Removing the smaller active spaces from \cref{fig:total_ccsd_best} significantly reduces the deviations for all UCC methods, indicating that increasing the active space size generally improves the agreement with CCSD. Despite the improvement in the performance of all four methods with the larger active spaces, the MAD for oo-UCCSD ($8.74$ ppm) and oo-fUCCSD ($8.62$ ppm) are still significantly better than for UCCSD ($50.44$ ppm) and fUCCSD ($50.46$ ppm).

Excluding the smaller active spaces substantially reduces the number of outliers; however, a few notable cases remain.
For oo-UCCSD and oo-fUCCSD the only outlier left is $\sigma_N$ in HCN(8,8) ($-95.16$ and $95.14$ ppm, respectively). Despite the improvement, UCCSD still has four shieldings that deviate by more than $50$ ppm from CCSD: $\sigma_\mathrm{O}$ and $\sigma_\mathrm{C}$ in CO(8,7) ($126.16$ and $133.68$ ppm, respectively), $\sigma_\mathrm{C}$ in HCN(8,8) ($95.03$ ppm) and $\sigma_\mathrm{N}$ in N$_2$(8,8) ($120.16$ ppm). The same four shieldings are outliers for fUCCSD. All the outliers continue to be shieldings in molecules with triple bonds.

In the comparison to CCSD, the factorized methods continue to be nearly indistinguishable from their non-factorized counterparts. This is evident both in the outliers discussed above and in the MD, MAD and SD where the difference is less than $0.5$ ppm between oo-UCCSD and oo-fUCCSD and between UCCSD and fUCCSD.

The shielding constants are separated in the deviations from CCSD arising from the diamagnetic and paramagnetic contribution in \cref{fig:dia_ccsd,fig:para_ccsd} in the Supporting Information. The figures reveal that the diamagnetic contribution is described well by all methods and active spaces and the error is exclusively arising from the description of the paramagnetic contribution.

Overall, oo-UCCSD and oo-fUCCSD show substantially better agreement with CCSD than their non-orbital-optimized counterparts, both across all active spaces and when only the largest active spaces are considered, highlighting the importance of orbital optimization for obtaining accurate NMR shielding constants.
Increasing the size of the active space, improves the agreement with CCSD for all four UCC methods, but does not eliminate the need for orbital optimization.

\subsection{Deviation from CCSD(T)}

\begin{figure}[h!]
    \centering
    \subfloat[all active spaces]{\includegraphics[width=0.85\linewidth]{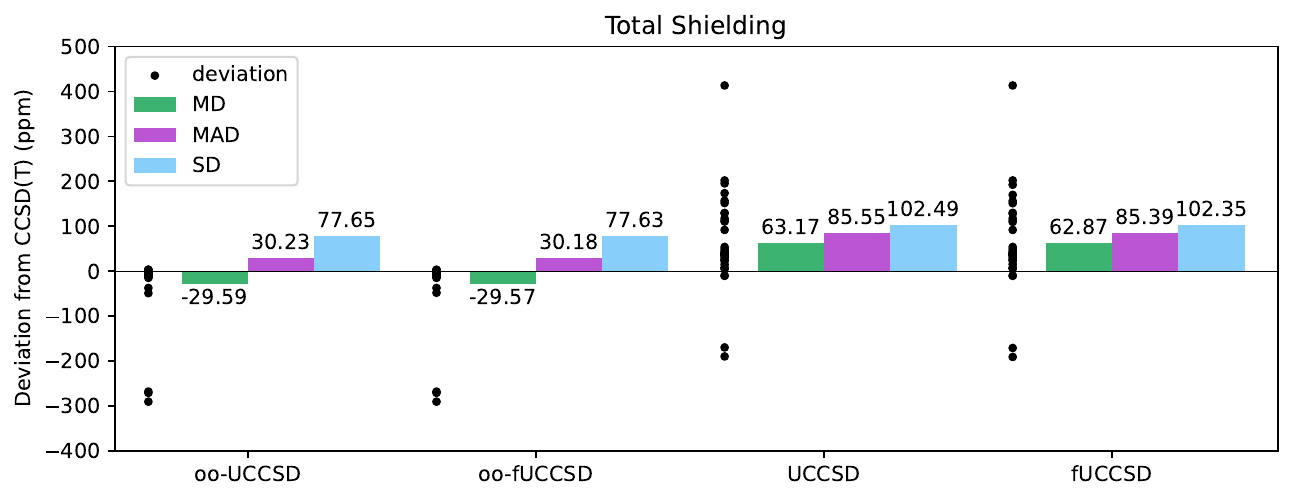}\label{fig:total_ccsd(t)_all}}
    
    \subfloat[largest active spaces]{\includegraphics[width=0.85\linewidth]{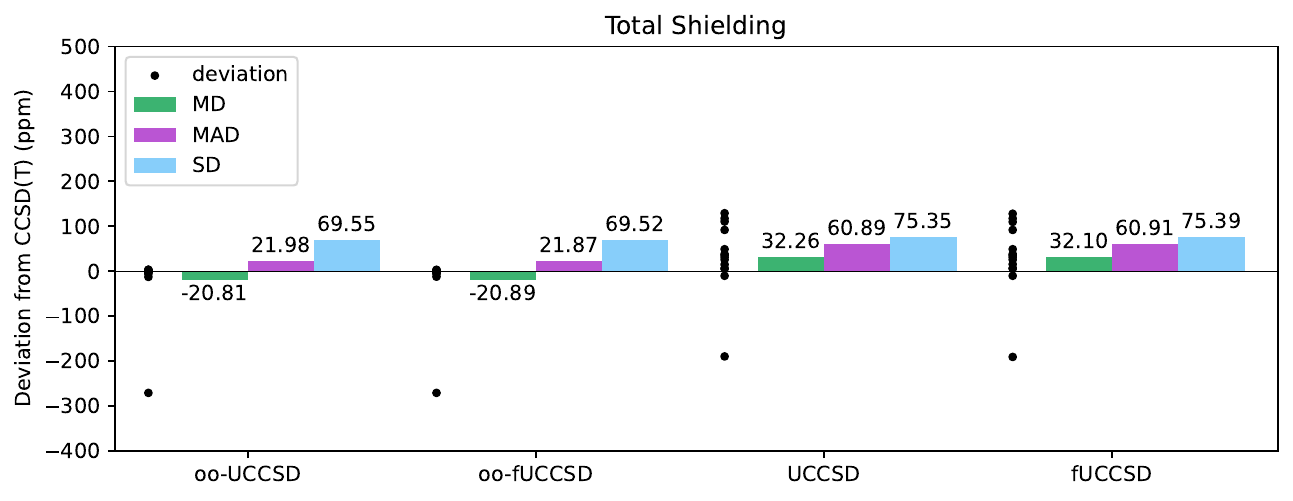}\label{fig:total_ccsd(t)_best}}
    \caption{Total shielding deviations, MD, MAD and SD of oo-UCCSD, oo-fUCCSD, UCCSD and fUCCSD from (full-space) CCSD(T). The points depict the individual deviations for every shielding in every molecule for \textbf{(a)} every active space (34 points) and \textbf{(b)} the largest active space (14 points). The bars depict the MD (green), MAD (purple), and SD (blue).}
    \label{fig:total_ccsd(t)}
\end{figure}

To further assess the impact of truncating the excitation operators at the doubles level in the UCC methods, the results were compared against CCSD(T). This comparison allows evaluation of how the deviations change when CCSD(T), rather than CCSD, is used as the reference. The deviations, together with the MD, MAD, and SD values for the four UCC methods relative to CCSD(T), are shown in \cref{fig:total_ccsd(t)} for both the complete set of active spaces and the largest active space of each molecule.

All shielding constants, except $\sigma_\mathrm{N}$ in HCN, change less than $10$ ppm going from CCSD to CCSD(T). The behavior for $\sigma_\mathrm{N}$ in HCN is unusual: while CCSD gives $187.83$ ppm, CCSD(T) gives $322.04$ ppm, nearly twice as big. 
This indicates strong correlation or complexity specific for $\sigma_\mathrm{N}$ in HCN, which requires further investigation, but does not change the overall conclusion drawn here about the performance of the UCC methods.
The inclusion of perturbative triples in the reference results in an increase in deviation across the board in \cref{fig:total_ccsd(t)_all,fig:total_ccsd(t)_best}, but the increase in, e.g., MAD is primarily due to $\sigma_\mathrm{N}$ in HCN, which is responsible for the outliers with significant negative deviations. Despite the increase in MAD, the orbital-optimized methods continue to deviate significantly less than the non-orbital-optimized methods for both all active spaces and only the largest active spaces.

Even with the inclusion of perturbative triples in the reference, the error for the UCC methods continues to arise exclusively from the paramagnetic contribution, which can be seen in \cref{fig:dia_ccsdpt,fig:para_ccsdpt} in the Supporting Information.

\section{Conclusion}

In this work, we report the calculation of NMR shielding constants in a quantum-computing-ready framework, to our knowledge the first of such applications. Our approach combines active space qLR with UCC ans{\"a}tze.

We employ the UCC and oo-UCC ans{\"a}tze, as well as their factorized counterparts, to compute NMR shielding constants for seven small molecules. Based on our calculations, we conclude that the factorization of the wavefunction has an insignificant impact on the shielding constants. The truncation of unitary coupled cluster after double excitations has a limited impact on the resulting shielding constants, although shielding constants for triple bonded atoms may require higher excitations. While the diamagnetic contribution is largely indifferent to the choice of wavefunction, the paramagnetic contribution varies significantly. Compared to full-space CCSD and CCSD(T), oo-UCC achieves a significantly better agreement than non-orbital-optimized UCC. Our results show that the calculation of NMR shielding constants on quantum hardware with oo-fUCC within a qLR framework is a promising approach: orbital optimization markedly improves the agreement with CCSD and CCSD(T) and factorization has an insignificant impact on the shielding constants.

Future work will focus on extending our approach for nuclear shielding constants to use gauge-including atomic orbitals to eliminate the gauge-origin dependence, and to investigate the effect of shot noise and device noise from quantum devices on the nuclear shielding constants.

\section*{Acknowledgments}
We acknowledge financial Support from the Novo Nordisk Foundation (NNF) for
the focused research project ``Hybrid Quantum Chemistry on
Hybrid Quantum Computers'' (NNF grant NNFSA220080996).

\section*{Supporting information}
Optimized geometries of all molecules, tabulated shielding constant results for all calculations, separation of all plots in the diamagnetic and paramagnetic contributions to the shielding constant.

\printbibliography \pagebreak

\include{SI_content}

\end{document}

%% file: SI_content.tex
\appendix
\counterwithin{figure}{section}
\counterwithin{table}{section}

\section{Coordinates of optimized structures in angstrom}
\subsection{Hydrogen molecule}
\begin{verbatim}
H 0.000000    0.000000    0.368761
H 0.000000    0.000000   -0.368761
\end{verbatim}

\subsection{Water}
\begin{verbatim}
O -0.000000   -0.000000    0.112729
H -0.000000    0.794937   -0.450915
H -0.000000   -0.794937   -0.450915
\end{verbatim}

\subsection{Ammonia}
\begin{verbatim}
N -0.000000    0.000000    0.074258
H  0.000000    0.979250   -0.173268
H -0.848055   -0.489625   -0.173268
H  0.848055   -0.489625   -0.173268
\end{verbatim}

\subsection{Methane}
\begin{verbatim}
C  0.000000    0.000000    0.000000
H  0.632771    0.632771    0.632771
H -0.632771   -0.632771    0.632771
H -0.632771    0.632771   -0.632771
H  0.632771   -0.632771   -0.632771
\end{verbatim}

\subsection{Carbon monoxide}
\begin{verbatim}
O  0.000000    0.000000    0.504281
C  0.000000    0.000000   -0.672374
\end{verbatim}

\subsection{Hydrogen fluoride}
\begin{verbatim}
F  0.000000    0.000000    0.094711
H  0.000000    0.000000   -0.852402
\end{verbatim}

\subsection{Hydrogen cyanide}
\begin{verbatim}
C  0.000000    0.000000   -0.522801
H  0.000000    0.000000   -1.593120
N  0.000000    0.000000    0.675704
\end{verbatim}

\subsection{Nitrogen molecule}
\begin{verbatim}
N  0.000000    0.000000    0.577226
N -0.000000    0.000000   -0.577226
\end{verbatim}
\clearpage

\section{Tabulated Shielding Constants} \label{sec:results_si}

\subsection{Water}

\begin{table}[h!]
    \caption{Hydrogen shielding constant, $\sigma_\mathrm{H}$, in H$_2$O for CASSCF, oo-UCCSD, oo-fUCCSD, CASCI, UCCSD, fUCCSD in three different active spaces as well as for full-space CCSD and CCSD(T) in the 6-31G** basis set (in ppm).} \label{tab:h2o_h}
    \centering 
    \begin{tabular}{rlccc}
        \toprule \multicolumn{2}{l}{H$_2$O, $\sigma_\mathrm{H}$} & Total & Diamagnetic & Paramagnetic \\ \midrule
        (4,4) & CASSCF & $56.70$ & $101.03$ & $-44.33$ \\ & oo-UCCSD & $56.69$ & $101.03$ & $-44.33$ \\ & oo-fUCCSD & $56.69$ & $101.03$ & $-44.33$ \\ & CASCI & $98.90$ & $100.90$ & $-2.00$ \\ & UCCSD & $98.90$ & $100.90$ & $-2.00$ \\ & fUCCSD & $98.90$ & $100.90$ & $-2.00$ \\ [1.0ex]
        (6,5) & CASSCF & $56.72$ & $101.03$ & $-44.31$ \\ & oo-UCCSD & $56.75$ & $101.04$ & $-44.29$ \\ & oo-fUCCSD & $56.75$ & $101.04$ & $-44.29$ \\ & CASCI & $97.83$ & $100.87$ & $-3.03$ \\ & UCCSD & $97.84$ & $100.87$ & $-3.03$ \\ & fUCCSD & $97.84$ & $100.87$ & $-3.03$ \\ [1.0ex]
        (6,9) & CASSCF & $56.88$ & $100.93$ & $-44.06$ \\ & oo-UCCSD & $56.84$ & $100.93$ & $-44.09$ \\ & oo-fUCCSD & $56.87$ & $100.93$ & $-44.06$ \\ & CASCI & $95.59$ & $101.02$ & $-5.43$ \\ & UCCSD & $95.60$ & $101.02$ & $-5.42$ \\ & fUCCSD & $95.60$ & $101.02$ & $-5.42$ \\ [1.0ex]
        (10,24) & CCSD & $57.67$ & $100.94$ & $-43.28$ \\ & CCSD(T) & $57.79$ & $100.96$ & $-43.17$ \\
    \bottomrule \end{tabular}
\end{table}

\begin{table}[h!]
    \caption{Oxygen shielding constant, $\sigma_\mathrm{O}$, in H$_2$O for CASSCF, oo-UCCSD, oo-fUCCSD, CASCI, UCCSD, fUCCSD in three different active spaces as well as for full-space CCSD and CCSD(T) in the 6-31G** basis set (in ppm).} \label{tab:h2o_o}
    \centering 
    \begin{tabular}{rlccc}
        \toprule \multicolumn{2}{l}{H$_2$O, $\sigma_\mathrm{O}$} & Total & Diamagnetic & Paramagnetic \\ \midrule
        (4,4) & CASSCF & $289.52$ & $415.68$ & $-126.16$ \\ & oo-UCCSD & $289.58$ & $415.68$ & $-126.10$ \\ & oo-fUCCSD & $289.57$ & $415.68$ & $-126.11$ \\ & CASCI & $337.22$ & $415.49$ & $-78.27$ \\ & UCCSD & $337.23$ & $415.49$ & $-78.26$ \\ & fUCCSD & $337.23$ & $415.49$ & $-78.26$ \\ [1.0ex]
        (6,5) & CASSCF & $289.98$ & $415.68$ & $-125.70$ \\ & oo-UCCSD & $289.25$ & $415.68$ & $-126.42$ \\ & oo-fUCCSD & $289.24$ & $415.67$ & $-126.43$ \\ & CASCI & $315.76$ & $415.24$ & $-99.48$ \\ & UCCSD & $316.15$ & $415.25$ & $-99.10$ \\ & fUCCSD & $316.15$ & $415.25$ & $-99.10$ \\ [1.0ex]
        (6,9) & CASSCF & $294.69$ & $415.64$ & $-120.95$ \\ & oo-UCCSD & $296.27$ & $415.64$ & $-119.37$ \\ & oo-fUCCSD & $294.93$ & $415.64$ & $-120.70$ \\ & CASCI & $298.92$ & $415.10$ & $-116.18$ \\ & UCCSD & $299.33$ & $415.10$ & $-115.77$ \\ & fUCCSD & $299.36$ & $415.10$ & $-115.75$ \\ [1.0ex]
        (10,24) & CCSD & $292.72$ & $415.21$ & $-122.49$ \\ & CCSD(T) & $293.26$ & $415.23$ & $-121.97$ \\
    \bottomrule \end{tabular}
\end{table}

\clearpage
\subsection{Methane}

\begin{table}[h!]
    \caption{Carbon shielding constant, $\sigma_\mathrm{C}$, in CH$_4$ for CASSCF, oo-UCCSD, oo-fUCCSD, CASCI, UCCSD, fUCCSD in three different active spaces as well as for full-space CCSD and CCSD(T) in the 6-31G** basis set (in ppm).} \label{tab:ch4_c}
    \centering 
    \begin{tabular}{rlccc}
        \toprule \multicolumn{2}{l}{CH$_4$, $\sigma_\mathrm{C}$} & Total & Diamagnetic & Paramagnetic \\ \midrule
        (6,7) & CASSCF & $211.18$ & $296.32$ & $-85.15$ \\ & oo-UCCSD & $210.29$ & $296.41$ & $-86.12$ \\ & oo-fUCCSD & $211.20$ & $296.32$ & $-85.13$ \\ & CASCI & $200.33$ & $295.83$ & $-95.49$ \\ & UCCSD & $200.33$ & $295.83$ & $-95.50$ \\ & fUCCSD & $200.33$ & $295.83$ & $-95.50$ \\ [1.0ex]
        (8,8) & CASSCF & $213.00$ & $296.50$ & $-83.50$ \\ & oo-UCCSD & $212.88$ & $296.50$ & $-83.62$ \\ & oo-fUCCSD & $212.86$ & $296.50$ & $-83.64$ \\ & CASCI & $199.79$ & $295.71$ & $-95.92$ \\ & UCCSD & $199.78$ & $295.71$ & $-95.94$ \\ & fUCCSD & $199.78$ & $295.72$ & $-95.93$ \\ [1.0ex]
        (10,34) & CCSD & $210.29$ & $296.49$ & $-86.20$ \\ & CCSD(T) & $210.27$ & $296.44$ & $-86.17$ \\
    \bottomrule \end{tabular}
\end{table}

\begin{table}[h!]
    \caption{Hydrogen shielding constant, $\sigma_\mathrm{H}$, in CH$_4$ for CASSCF, oo-UCCSD, oo-fUCCSD, CASCI, UCCSD, fUCCSD in three different active spaces as well as for full-space CCSD and CCSD(T) in the 6-31G** basis set (in ppm).} \label{tab:h2_hh}
    \centering 
    \begin{tabular}{rlccc}
        \toprule \multicolumn{2}{l}{CH$_4$, $\sigma_\mathrm{H}$} & Total & Diamagnetic & Paramagnetic \\ \midrule
        (6,7) & CASSCF & $50.77$ & $87.22$ & $-36.45$ \\ & oo-UCCSD & $50.72$ & $87.18$ & $-36.46$ \\ & oo-fUCCSD & $50.79$ & $87.18$ & $-36.39$ \\ & CASCI & $76.61$ & $87.09$ & $-10.48$ \\ & UCCSD & $76.61$ & $87.10$ & $-10.48$ \\ & fUCCSD & $76.61$ & $87.10$ & $-10.48$ \\ [1.0ex]
        (8,8) & CASSCF & $50.72$ & $87.18$ & $-36.45$ \\ & oo-UCCSD & $50.73$ & $87.18$ & $-36.45$ \\ & oo-fUCCSD & $50.73$ & $87.18$ & $-36.44$ \\ & CASCI & $76.62$ & $87.06$ & $-10.45$ \\ & UCCSD & $76.62$ & $87.07$ & $-10.45$ \\ & fUCCSD & $76.62$ & $87.07$ & $-10.45$ \\ [1.0ex]
        (10,34) & CCSD & $51.15$ & $87.20$ & $-36.05$ \\ & CCSD(T) & $51.20$ & $87.19$ & $-35.99$ \\
    \bottomrule \end{tabular}
\end{table}

\clearpage
\subsection{Ammonia}

\begin{table}[h!]
    \caption{Nitrogen shielding constant, $\sigma_\mathrm{N}$, in NH$_3$ for CASSCF, oo-UCCSD, oo-fUCCSD, CASCI, UCCSD, fUCCSD in three different active spaces as well as for full-space CCSD and CCSD(T) in the 6-31G** basis set (in ppm).} \label{tab:nh3_n}
    \centering 
    \begin{tabular}{rlccc}
        \toprule \multicolumn{2}{l}{NH$_3$, $\sigma_\mathrm{N}$} & Total & Diamagnetic & Paramagnetic \\ \midrule
        (6,6) & CASSCF & $262.35$ & $354.37$ & $-92.02$ \\ & oo-UCCSD & $262.31$ & $354.37$ & $-92.06$ \\ & oo-fUCCSD & $262.29$ & $354.37$ & $-92.07$ \\ & CASCI & $267.14$ & $353.80$ & $-86.66$ \\ & UCCSD & $267.26$ & $353.81$ & $-86.55$ \\ & fUCCSD & $267.26$ & $353.81$ & $-86.54$ \\ [1.0ex]
        (8,7) & CASSCF & $262.73$ & $354.37$ & $-91.64$ \\ & oo-UCCSD & $262.37$ & $354.37$ & $-92.00$ \\ & oo-fUCCSD & $262.35$ & $354.37$ & $-92.02$ \\ & CASCI & $265.36$ & $353.65$ & $-88.29$ \\ & UCCSD & $265.58$ & $353.66$ & $-88.08$ \\ & fUCCSD & $265.59$ & $353.66$ & $-88.07$ \\ [1.0ex]
        (10,29) & CCSD & $259.81$ & $354.24$ & $-94.44$ \\ & CCSD(T) & $259.76$ & $354.21$ & $-94.45$ \\
    \bottomrule \end{tabular}
\end{table}

\begin{table}[h!]
    \caption{Hydrogen shielding constant, $\sigma_\mathrm{H}$, in NH$_3$ for CASSCF, oo-UCCSD, oo-fUCCSD, CASCI, UCCSD, fUCCSD in three different active spaces as well as for full-space CCSD and CCSD(T) in the 6-31G** basis set (in ppm).} \label{tab:nh3_h}
    \centering 
    \begin{tabular}{rlccc}
        \toprule \multicolumn{2}{l}{NH$_3$, $\sigma_\mathrm{H}$} & Total & Diamagnetic & Paramagnetic \\ \midrule
        (6,6) & CASSCF & $54.25$ & $95.19$ & $-40.94$ \\ & oo-UCCSD & $54.24$ & $95.19$ & $-40.96$ \\ & oo-fUCCSD & $54.24$ & $95.19$ & $-40.95$ \\ & CASCI & $90.22$ & $95.12$ & $-4.90$ \\ & UCCSD & $90.22$ & $95.12$ & $-4.90$ \\ & fUCCSD & $90.22$ & $95.12$ & $-4.90$ \\ [1.0ex]
        (8,7) & CASSCF & $54.27$ & $95.19$ & $-40.93$ \\ & oo-UCCSD & $54.30$ & $95.19$ & $-40.90$ \\ & oo-UCCSD & $54.30$ & $95.19$ & $-40.90$ \\ & CASCI & $90.33$ & $95.08$ & $-4.75$ \\ & UCCSD & $90.35$ & $95.09$ & $-4.74$ \\ & fUCCSD & $90.35$ & $95.09$ & $-4.74$ \\ [1.0ex]
        (10,29) & CCSD & $54.97$ & $95.21$ & $-40.23$ \\ & CCSD(T) & $55.07$ & $95.21$ & $-40.14$ \\
    \bottomrule \end{tabular}
\end{table}

\clearpage
\subsection{Carbon monoxide}

\begin{table}[h!]
    \caption{Oxygen shielding constant, $\sigma_\mathrm{O}$, in CO for CASSCF, oo-UCCSD, oo-fUCCSD, CASCI, UCCSD, fUCCSD in three different active spaces as well as for full-space CCSD and CCSD(T) in the 6-31G* basis set (in ppm).} \label{tab:co_o}
    \centering 
    \begin{tabular}{rlccc}
        \toprule \multicolumn{2}{l}{CO, $\sigma_\mathrm{O}$} & Total & Diamagnetic & Paramagnetic \\ \midrule
        (6,5) & CASSCF & $-74.13$ & $442.25$ & $-516.38$ \\ & oo-UCCSD & $-72.63$ & $442.22$ & $-514.85$ \\ & oo-fUCCSD & $-71.69$ & $442.23$ & $-513.92$ \\ & CASCI & $171.78$ & $441.20$ & $-269.42$ \\ & UCCSD & $168.79$ & $441.28$ & $-272.49$ \\ & fUCCSD & $168.99$ & $441.30$ & $-272.32$ \\ [1.0ex]
        (6,6) & CASSCF & $-36.97$ & $442.37$ & $-479.35$ \\ & oo-UCCSD & $-36.57$ & $442.36$ & $-478.93$ \\ & oo-fUCCSD & $-36.35$ & $442.37$ & $-478.72$ \\ & CASCI & $150.02$ & $441.19$ & $-291.17$ \\ & UCCSD & $146.31$ & $441.26$ & $-294.95$ \\ & fUCCSD & $146.51$ & $441.28$ & $-294.78$ \\ [1.0ex]
        (8,7) & CASSCF & $-41.13$ & $442.39$ & $-483.53$ \\ & oo-UCCSD & $-36.49$ & $442.36$ & $-478.85$ \\ & oo-fUCCSD & $-36.01$ & $442.37$ & $-478.39$ \\ & CASCI & $93.77$ & $441.26$ & $-347.49$ \\ & UCCSD & $93.74$ & $441.34$ & $-347.60$ \\ & fUCCSD & $93.90$ & $441.36$ & $-347.46$ \\ [1.0ex]
        (14,28) & CCSD & $-32.39$ & $441.86$ & $-474.24$ \\ & CCSD(T) & $-23.46$ & $441.87$ & $-465.33$ \\
    \bottomrule \end{tabular}
\end{table}

\begin{table}[h!]
    \caption{Carbon shielding constant, $\sigma_\mathrm{C}$, in CO for CASSCF, oo-UCCSD, oo-fUCCSD, CASCI, UCCSD, fUCCSD in three different active spaces as well as for full-space CCSD and CCSD(T) in the 6-31G* basis set (in ppm).} \label{tab:co_c}
    \centering 
    \begin{tabular}{rlccc}
        \toprule \multicolumn{2}{l}{CO, $\sigma_\mathrm{C}$} & Total & Diamagnetic & Paramagnetic \\ \midrule
        (6,5) & CASSCF & $23.91$ & $324.03$ & $-300.12$ \\ & oo-UCCSD & $24.18$ & $323.97$ & $-299.80$ \\ & oo-fUCCSD & $24.06$ & $323.97$ & $-299.91$ \\ & CASCI & $190.62$ & $323.67$ & $-133.05$ \\ & UCCSD & $190.20$ & $323.66$ & $-133.46$ \\ & fUCCSD & $190.16$ & $323.65$ & $-133.50$ \\ [1.0ex]
        (6,6) & CASSCF & $29.80$ & $324.17$ & $-294.37$ \\ & oo-UCCSD & $28.56$ & $324.13$ & $-295.57$ \\ & oo-fUCCSD & $28.41$ & $324.13$ & $-295.72$ \\ & CASCI & $195.72$ & $323.69$ & $-127.97$ \\ & UCCSD & $195.18$ & $323.68$ & $-128.50$ \\ & fUCCSD & $195.12$ & $323.67$ & $-128.55$ \\ [1.0ex]
        (8,7) & CASSCF & $32.87$ & $324.27$ & $-291.40$ \\ & oo-UCCSD & $32.77$ & $324.21$ & $-291.44$ \\ & oo-fUCCSD & $32.58$ & $324.20$ & $-291.62$ \\ & CASCI & $168.02$ & $323.65$ & $-155.63$ \\ & UCCSD & $166.87$ & $323.65$ & $-156.77$ \\ & fUCCSD & $166.75$ & $323.64$ & $-156.89$ \\ [1.0ex]
        (14,28) & CCSD & $34.34$ & $324.31$ & $-289.97$ \\ & CCSD(T) & $39.16$ & $324.33$ & $-285.17$ \\
    \bottomrule \end{tabular}
\end{table}

\clearpage
\subsection{Hydrogen fluoride}

\begin{table}[h!]
    \caption{Hydrogen shielding constant, $\sigma_\mathrm{H}$, in HF for CASSCF, oo-UCCSD, oo-fUCCSD, CASCI, UCCSD, fUCCSD in (6,7) as well as for full-space CCSD and CCSD(T) in the 6-31G** basis set (in ppm).} \label{tab:hf_h}
    \centering 
    \begin{tabular}{rlccc}
        \toprule \multicolumn{2}{l}{HF, $\sigma_\mathrm{H}$} & Total & Diamagnetic & Paramagnetic \\ \midrule
        (6,7) & CASSCF & $60.03$ & $105.91$ & $-45.88$ \\ & oo-UCCSD & $59.99$ & $105.90$ & $-45.91$ \\ & oo-fUCCSD & $59.98$ & $105.90$ & $-45.92$ \\ & CASCI & $109.64$ & $106.16$ & $3.48$ \\ & UCCSD & $109.64$ & $106.16$ & $3.48$ \\ & fUCCSD & $109.64$ & $106.16$ & $3.48$ \\ [1.0ex]
        (10,19) & CCSD & $60.84$ & $105.87$ & $-45.02$ \\ & CCSD(T) & $60.90$ & $105.88$ & $-44.98$ \\
    \bottomrule \end{tabular}
\end{table}

\begin{table}[h!]
    \caption{Fluorine shielding constant, $\sigma_\mathrm{F}$, in HF for CASSCF, oo-UCCSD, oo-fUCCSD, CASCI, UCCSD, fUCCSD in (6,7) as well as for full-space CCSD and CCSD(T) in the 6-31G** basis set (in ppm).} \label{tab:hf_f}
    \centering 
    \begin{tabular}{rlccc}
        \toprule \multicolumn{2}{l}{HF, $\sigma_\mathrm{F}$} & Total & Diamagnetic & Paramagnetic \\ \midrule
        (6,7) & CASSCF & $361.08$ & $481.90$ & $-120.82$ \\ & oo-UCCSD & $361.33$ & $481.90$ & $-120.56$ \\ & oo-fUCCSD & $361.36$ & $481.90$ & $-120.54$ \\ & CASCI & $377.51$ & $481.52$ & $-104.01$ \\ & UCCSD & $378.04$ & $481.51$ & $-103.47$ \\ & fUCCSD & $378.10$ & $481.52$ & $-103.42$ \\ [1.0ex]
        (10,19) & CCSD & $360.99$ & $481.24$ & $-120.25$ \\ & CCSD(T) & $362.84$ & $481.33$ & $-118.49$ \\
    \bottomrule \end{tabular}
\end{table}

\clearpage
\subsection{Hydrogen cyanide}

\begin{table}[h!]
    \caption{Hydrogen shielding constant, $\sigma_\mathrm{H}$, in HCN for CASSCF, oo-UCCSD, oo-fUCCSD, CASCI, UCCSD, fUCCSD in three different active spaces as well as for full-space CCSD and CCSD(T) in the 6-31G** basis set (in ppm).} \label{tab:hcn_h}
    \centering 
    \begin{tabular}{rlccc}
        \toprule \multicolumn{2}{l}{HCN, $\sigma_\mathrm{H}$} & Total & Diamagnetic & Paramagnetic \\ \midrule
        (4,4) & CASSCF & $61.03$ & $98.93$ & $-37.90$ \\ & oo-UCCSD & $61.03$ & $98.93$ & $-37.90$ \\ & oo-fUCCSD & $61.03$ & $98.93$ & $-37.89$ \\ & CASCI & $98.82$ & $98.82$ & $0.00$ \\ & UCCSD & $98.83$ & $98.83$ & $0.00$ \\ & fUCCSD & $98.83$ & $98.83$ & $0.00$ \\ [1.0ex]
        (6,6) & CASSCF & $61.19$ & $99.04$ & $-37.85$ \\ & oo-UCCSD & $61.20$ & $99.03$ & $-37.84$ \\ & oo-fUCCSD & $61.20$ & $99.03$ & $-37.84$ \\ & CASCI & $98.84$ & $98.87$ & $-0.03$ \\ & UCCSD & $98.87$ & $98.87$ & $0.00$ \\ & fUCCSD & $98.87$ & $98.87$ & $0.00$ \\ [1.0ex]
        (8,8) & CASSCF & $61.04$ & $99.01$ & $-37.96$ \\ & oo-UCCSD & $61.47$ & $99.23$ & $-37.77$ \\ & oo-fUCCSD & $61.47$ & $99.23$ & $-37.76$ \\ & CASCI & $92.74$ & $98.64$ & $-5.90$ \\ & UCCSD & $92.78$ & $98.65$ & $-5.87$ \\ & fUCCSD & $92.77$ & $98.64$ & $-5.87$ \\ [1.0ex]
        (14,28) & CCSD & $61.87$ & $99.11$ & $-37.24$ \\ & CCSD(T) & $61.97$ & $99.13$ & $-37.16$ \\
    \bottomrule \end{tabular}
\end{table}

\begin{table}[h!]
    \caption{Carbon shielding constant, $\sigma_\mathrm{C}$, in HCN for CASSCF, oo-UCCSD, oo-fUCCSD, CASCI, UCCSD, fUCCSD in three different active spaces as well as for full-space CCSD and CCSD(T) in the 6-31G** basis set (in ppm).} \label{tab:hcn_c}
    \centering 
    \begin{tabular}{rlccc}
        \toprule \multicolumn{2}{l}{HCN, $\sigma_\mathrm{C}$} & Total & Diamagnetic & Paramagnetic \\ \midrule
        (4,4) & CASSCF & $114.70$ & $324.48$ & $-209.78$ \\ & oo-UCCSD & $114.49$ & $324.48$ & $-209.98$ \\ & oo-fUCCSD & $114.48$ & $324.48$ & $-210.00$ \\ & CASCI & $323.90$ & $323.90$ & $0.00$ \\ & UCCSD & $323.92$ & $323.92$ & $0.00$ \\ & fUCCSD & $323.92$ & $323.92$ & $0.00$ \\ [1.0ex]
        (6,6) & CASSCF & $119.13$ & $324.60$ & $-205.47$ \\ & oo-UCCSD & $119.63$ & $324.60$ & $-204.96$ \\ & oo-fUCCSD & $119.62$ & $324.60$ & $-204.98$ \\ & CASCI & $233.95$ & $323.90$ & $-89.94$ \\ & UCCSD & $233.72$ & $323.92$ & $-90.20$ \\ & fUCCSD & $233.67$ & $323.92$ & $-90.24$ \\ [1.0ex]
        (8,8) & CASSCF & $118.48$ & $324.54$ & $-206.06$ \\ & oo-UCCSD & $119.42$ & $324.72$ & $-205.29$ \\ & oo-fUCCSD & $119.41$ & $324.71$ & $-205.31$ \\ & CASCI & $213.69$ & $323.85$ & $-110.17$ \\ & UCCSD & $213.63$ & $323.88$ & $-110.24$ \\ & fUCCSD & $213.58$ & $323.87$ & $-110.29$ \\ [1.0ex]
        (14,28) & CCSD & $118.65$ & $324.72$ & $-206.07$ \\ & CCSD(T) & $122.05$ & $324.71$ & $-202.67$ \\
    \bottomrule \end{tabular}
\end{table}

\begin{table}[h!]
    \caption{Nitrogen shielding constant, $\sigma_\mathrm{N}$, in HCN for CASSCF, oo-UCCSD, oo-fUCCSD, CASCI, UCCSD, fUCCSD in three different active spaces as well as for full-space CCSD and CCSD(T) in the 6-31G** basis set (in ppm).} \label{tab:hcn_n}
    \centering 
    \begin{tabular}{rlccc}
        \toprule \multicolumn{2}{l}{HCN, $\sigma_\mathrm{N}$} & Total & Diamagnetic & Paramagnetic \\ \midrule
        (4,4) & CASSCF & $31.86$ & $376.75$ & $-344.89$ \\ & oo-UCCSD & $31.39$ & $376.74$ & $-345.36$ \\ & oo-fUCCSD & $31.40$ & $376.74$ & $-345.35$ \\ & CASCI & $375.84$ & $375.84$ & $0.00$ \\ & UCCSD & $375.82$ & $375.82$ & $0.00$ \\ & fUCCSD & $375.82$ & $375.82$ & $0.00$ \\ [1.0ex]
        (6,6) & CASSCF & $51.41$ & $376.96$ & $-325.55$ \\ & oo-UCCSD & $53.81$ & $376.95$ & $-323.14$ \\ & oo-fUCCSD & $53.79$ & $376.95$ & $-323.16$ \\ & CASCI & $152.06$ & $375.75$ & $-223.69$ \\ & UCCSD & $150.85$ & $375.74$ & $-224.89$ \\ & fUCCSD & $150.87$ & $375.75$ & $-224.87$ \\ [1.0ex]
        (8,8) & CASSCF & $57.52$ & $376.90$ & $-319.39$ \\ & oo-UCCSD & $50.85$ & $376.94$ & $-326.09$ \\ & oo-fUCCSD & $50.87$ & $376.94$ & $-326.07$ \\ & CASCI & $132.05$ & $375.74$ & $-243.69$ \\ & UCCSD & $130.90$ & $375.72$ & $-244.83$ \\ & fUCCSD & $130.93$ & $375.73$ & $-244.80$ \\ [1.0ex]
        (14,28) & CCSD & $146.01$ & $375.94$ & $-229.93$ \\ & CCSD(T) & $322.04$ & $376.02$ & $-53.98$ \\
    \bottomrule \end{tabular}
\end{table}

\clearpage
\subsection{Dinitrogen}

\begin{table}[h!]
    \caption{Nitrogen shielding constant, $\sigma_\mathrm{N}$, in N$_2$ for CASSCF, oo-UCCSD, oo-fUCCSD, CASCI, UCCSD, fUCCSD in three different active spaces as well as for full-space CCSD and CCSD(T) in the 6-31G* basis set (in ppm).} \label{tab:n2_n}
    \centering 
    \begin{tabular}{rlccc}
        \toprule \multicolumn{2}{l}{N$_2$, $\sigma_\mathrm{N}$} & Total & Diamagnetic & Paramagnetic \\ \midrule
        (4,4) & CASSCF & $-69.91$ & $381.16$ & $-451.07$ \\ & oo-UCCSD & $-70.52$ & $381.15$ & $-451.67$ \\ & oo-fUCCSD & $-70.52$ & $381.15$ & $-451.67$ \\ & CASCI & $380.37$ & $380.37$ & $0.00$ \\ & UCCSD & $380.37$ & $380.37$ & $0.00$ \\ & fUCCSD & $380.37$ & $380.37$ & $0.00$ \\ [1.0ex]
        (6,6) & CASSCF & $-45.72$ & $381.37$ & $-427.08$ \\ & oo-UCCSD & $-43.07$ & $381.36$ & $-424.43$ \\ & oo-fUCCSD & $-43.09$ & $381.36$ & $-424.45$ \\ & CASCI & $96.49$ & $380.36$ & $-283.87$ \\ & UCCSD & $96.94$ & $380.36$ & $-283.41$ \\ & fUCCSD & $96.94$ & $380.36$ & $-283.42$ \\ [1.0ex]
        (8,8) & CASSCF & $-39.88$ & $381.38$ & $-421.25$ \\ & oo-UCCSD & $-34.10$ & $381.37$ & $-415.47$ \\ & oo-fUCCSD & $-34.15$ & $381.37$ & $-415.52$ \\ & CASCI & $77.51$ & $380.38$ & $-302.86$ \\ & UCCSD & $76.56$ & $380.38$ & $-303.82$ \\ & fUCCSD & $76.56$ & $380.38$ & $-303.83$ \\ [1.0ex]
        (14,28) & CCSD & $-42.65$ & $381.21$ & $-423.86$ \\ & CCSD(T) & $-32.90$ & $381.18$ & $-414.08$ \\
    \bottomrule \end{tabular}
\end{table}

\clearpage

\section{Breakdown of Shielding Constants in Diamagnetic and Paramagnetic Contributions}

\subsection{Deviations from CASSCF and CASCI}

\begin{figure}[h!]
    \centering
    \subfloat[with orbital rotations]{\includegraphics[width=0.4\linewidth]{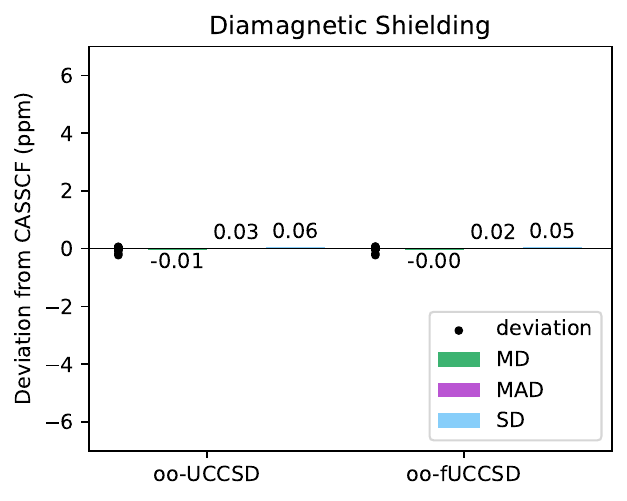}\label{fig:dia_casscf}}
    \subfloat[without orbital rotations]{\includegraphics[width=0.4\linewidth]{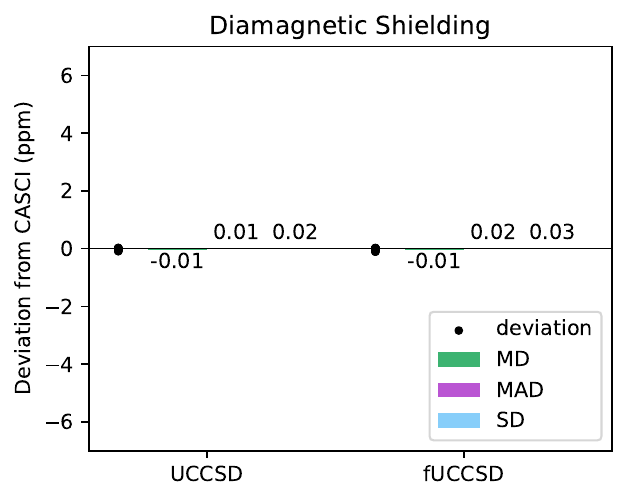}\label{fig:dia_casci}}
    \caption{Diamagnetic shielding deviations, MD, MAD and SD of \textbf{(a)} oo-UCCSD and oo-fUCCSD from CASSCF and \textbf{(b)} UCCSD and fUCCSD from CASCI. The 34 points depict the individual deviations for every shielding in every molecule and active space. The bars depict the MD (green), MAD (purple), and SD (blue).}
    \label{fig:dia_both}
\end{figure}

\begin{figure}[h!]
    \centering
    \subfloat[with orbital rotations]{\includegraphics[width=0.4\linewidth]{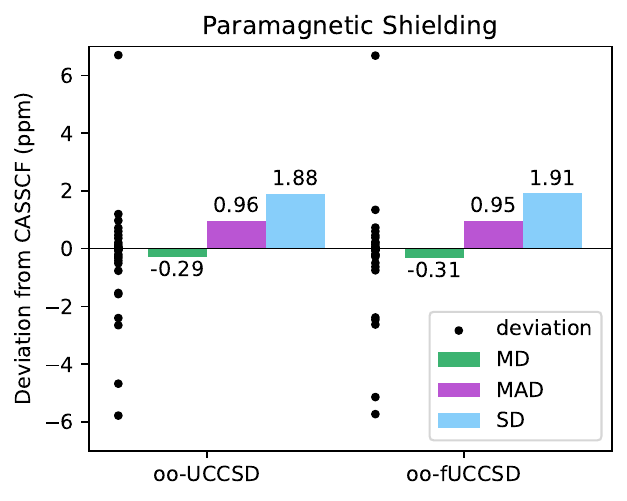}\label{fig:para_casscf}}
    \subfloat[without orbital rotations]{\includegraphics[width=0.4\linewidth]{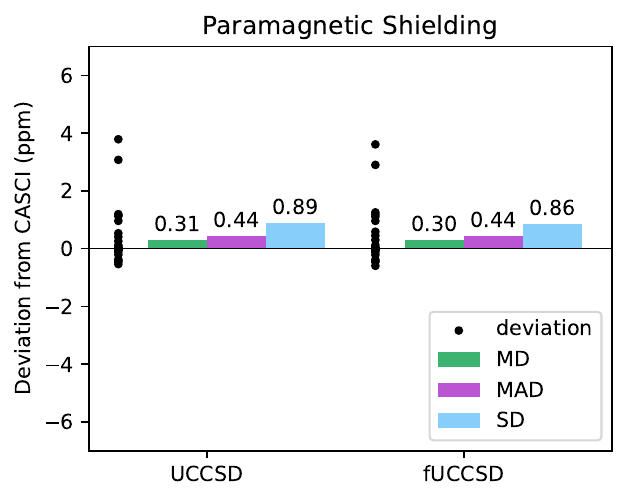}\label{fig:para_casci}}
    \caption{Paramagnetic shielding deviations, MD, MAD and SD of \textbf{(a)} oo-UCCSD and oo-fUCCSD from CASSCF and \textbf{(b)} UCCSD and fUCCSD from CASCI. The 34 points depict the individual deviations for every shielding in every molecule and active space. The bars depict the MD (green), MAD (purple), and SD (blue).}
    \label{fig:para_both}
\end{figure}

\clearpage
\subsection{Deviations from CCSD}

\begin{figure}[h!]
    \centering
    \subfloat[all active spaces]{\includegraphics[width=0.85\linewidth]{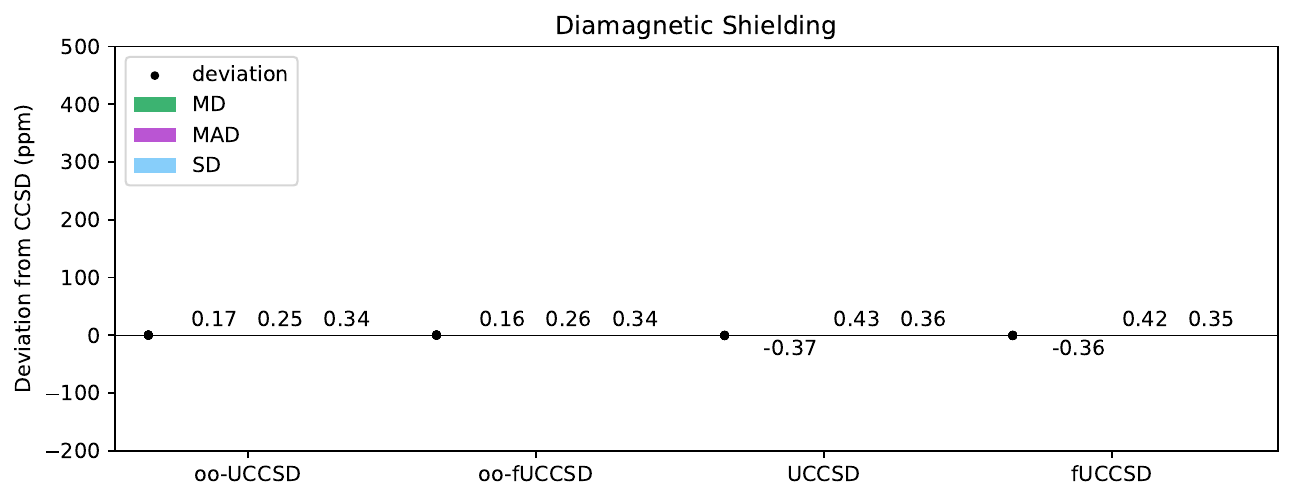}\label{fig:dia_ccsd_all}}
    
    \subfloat[largest active spaces]{\includegraphics[width=0.85\linewidth]{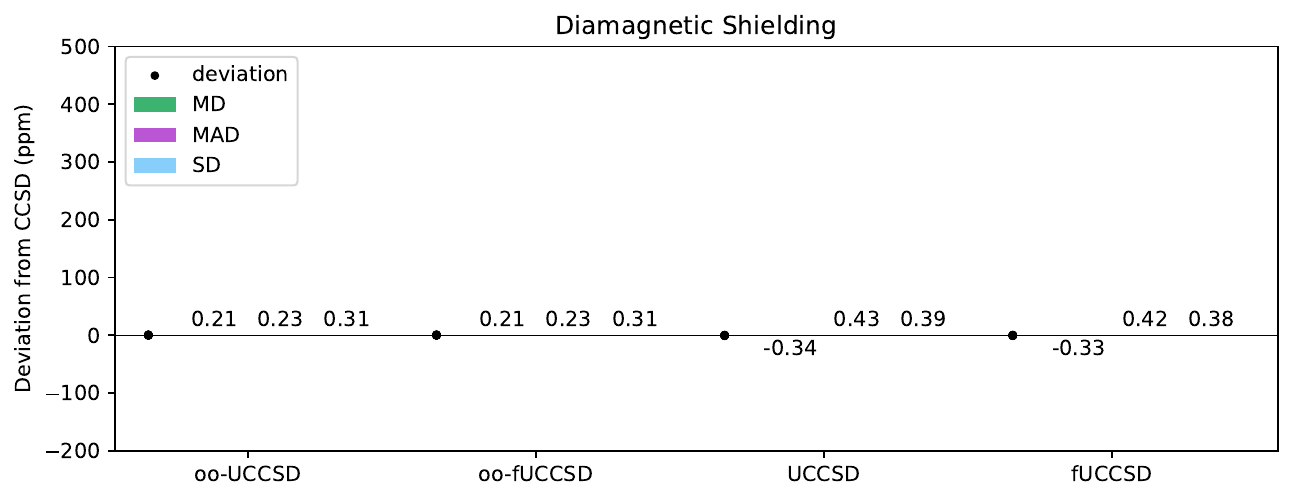}\label{fig:dia_ccsd_best}}
    \caption{Diamagnetic shielding deviations, MD, MAD and SD of oo-UCCSD, oo-fUCCSD, UCCSD and fUCCSD from (full-space) CCSD. The points depict the individual deviations for every shielding in every molecule for \textbf{(a)} every active space (34 points) and \textbf{(b)} the largest active space (14 points). The bars depict the MD (green), MAD (purple), and SD (blue).}
    \label{fig:dia_ccsd}
\end{figure}

\begin{figure}[h!]
    \centering
    \subfloat[all active spaces]{\includegraphics[width=0.85\linewidth]{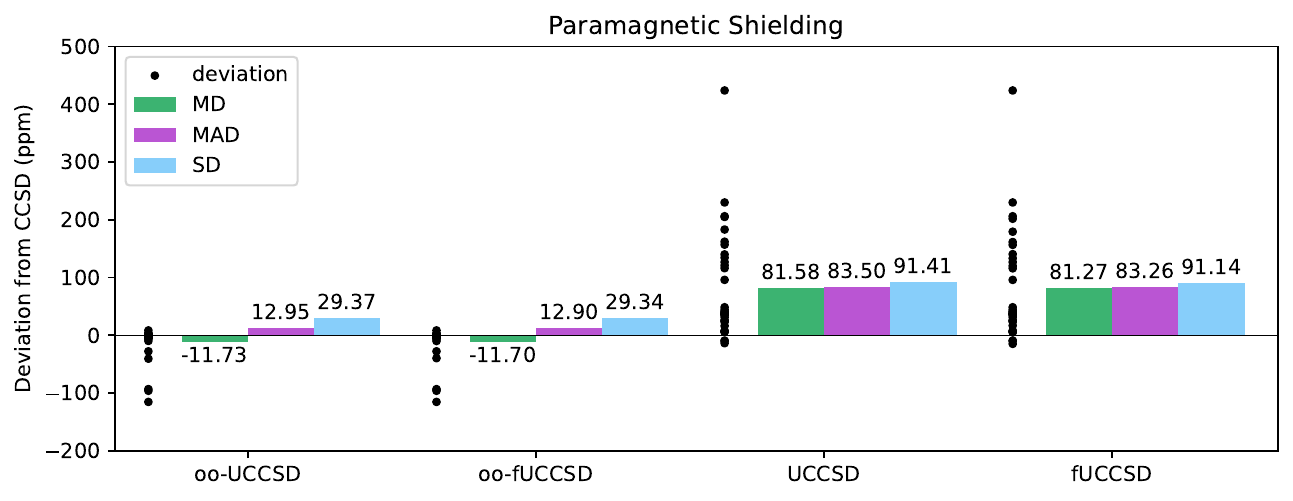}\label{fig:para_ccsd_all}}
    
    \subfloat[largest active spaces]{\includegraphics[width=0.85\linewidth]{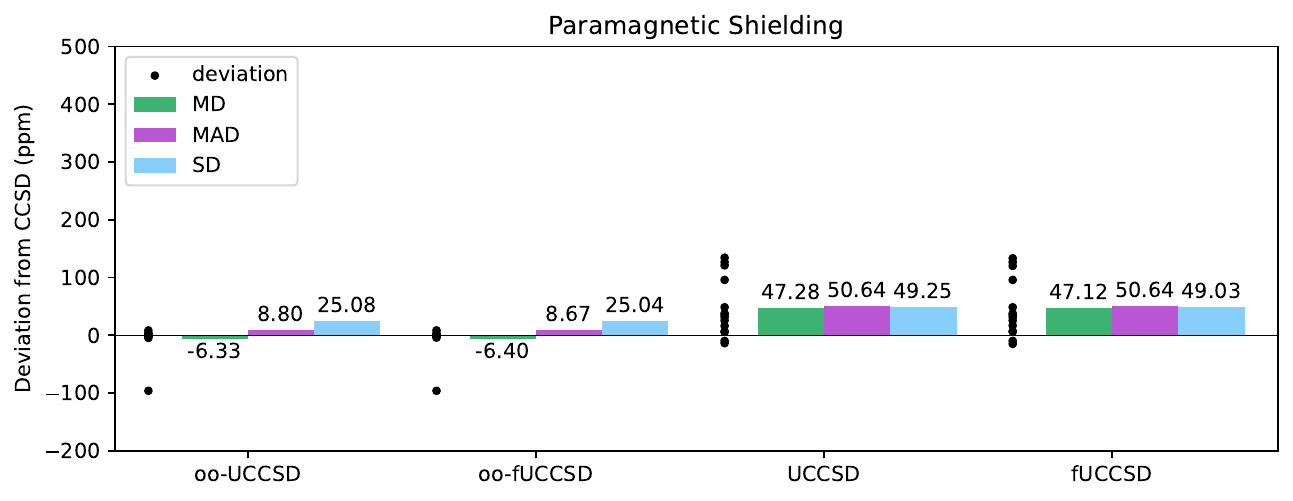}\label{fig:para_ccsd_best}}
    \caption{Paramagnetic shielding deviations, MD, MAD and SD of oo-UCCSD, oo-fUCCSD, UCCSD and fUCCSD from (full-space) CCSD. The points depict the individual deviations for every shielding in every molecule for \textbf{(a)} every active space (34 points) and \textbf{(b)} the largest active space (14 points). The bars depict the MD (green), MAD (purple), and SD (blue).}
    \label{fig:para_ccsd}
\end{figure}

\clearpage
\subsection{Deviations from CCSD(T)}

\begin{figure}[h!]
    \centering
    \subfloat[all active spaces]{\includegraphics[width=0.85\linewidth]{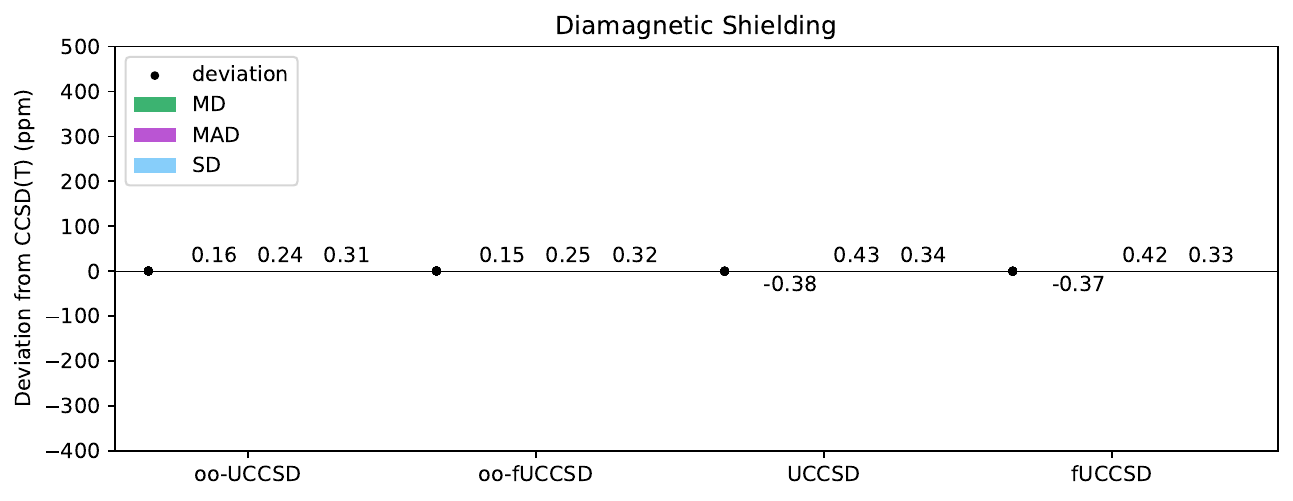}\label{fig:dia_ccsdpt_all}}
    
    \subfloat[largest active spaces]{\includegraphics[width=0.85\linewidth]{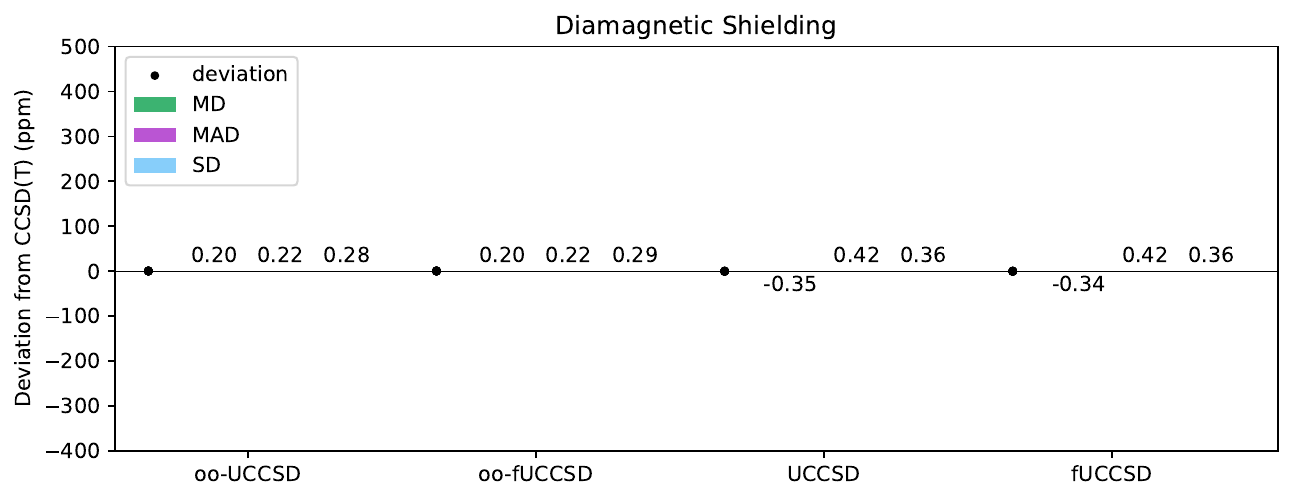}\label{fig:dia_ccsdpt_best}}
    \caption{Diamagnetic shielding deviations, MD, MAD and SD of oo-UCCSD, oo-fUCCSD, UCCSD and fUCCSD from (full-space) CCSD(T). The points depict the individual deviations for every shielding in every molecule for \textbf{(a)} every active space (34 points) and \textbf{(b)} the largest active space (14 points). The bars depict the MD (green), MAD (purple), and SD (blue).}
    \label{fig:dia_ccsdpt}
\end{figure}

\begin{figure}[h!]
    \centering
    \subfloat[all active spaces]{\includegraphics[width=0.85\linewidth]{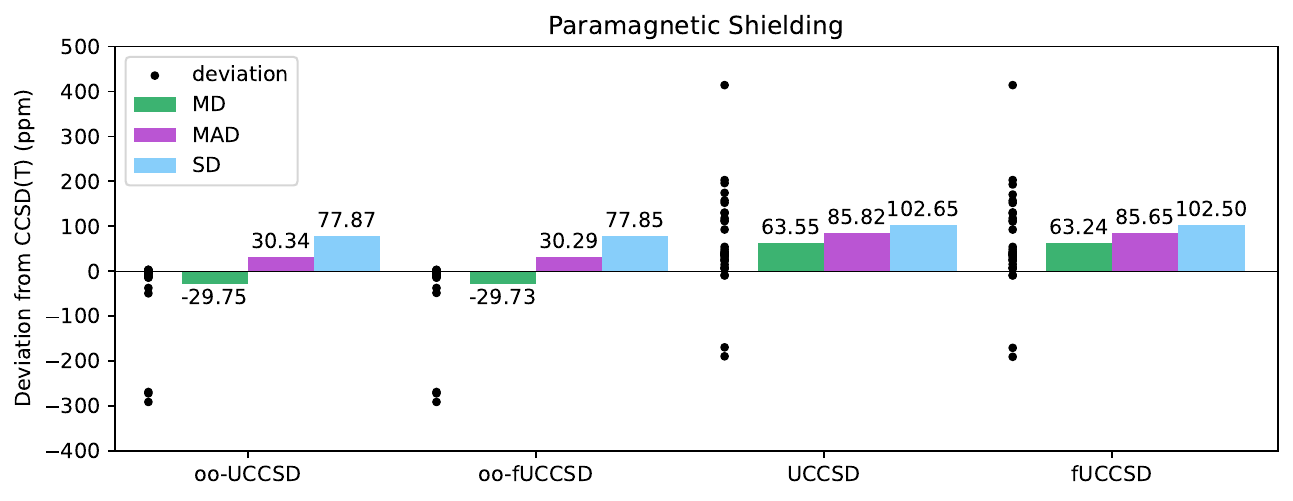}\label{fig:para_ccsdpt_all}}
    
    \subfloat[largest active spaces]{\includegraphics[width=0.85\linewidth]{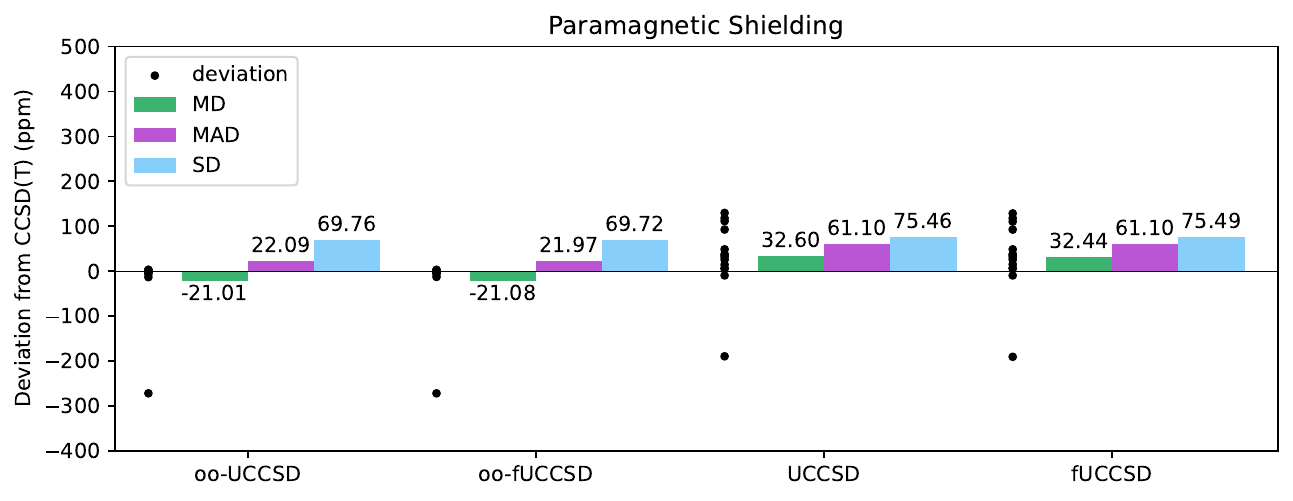}\label{fig:para_ccsdpt_best}}
    \caption{Paramagnetic shielding deviations, MD, MAD and SD of oo-UCCSD, oo-fUCCSD, UCCSD and fUCCSD from (full-space) CCSD(T). The points depict the individual deviations for every shielding in every molecule for \textbf{(a)} every active space (34 points) and \textbf{(b)} the largest active space (14 points). The bars depict the MD (green), MAD (purple), and SD (blue).}
    \label{fig:para_ccsdpt}
\end{figure}

\clearpage